\documentclass[showpacs,superscriptaddress,reprint,nofootinbib,amsmath,amssymb,aps,pre]{revtex4-1}
\usepackage{amssymb}
\usepackage{color}
\usepackage{graphicx}
\usepackage{amsmath}
\usepackage{mathrsfs}
\usepackage{times}
\usepackage{subeqnarray}
\usepackage{cases}
\usepackage{bm}
\usepackage{subfigure}
\usepackage{dcolumn}
\usepackage{hyperref}
\usepackage{outlines}
\usepackage{enumitem}
\setenumerate[1]{label=\Roman*.}   
\setenumerate[2]{label=\arabic*.}
\usepackage{booktabs}
\usepackage[table,xcdraw]{xcolor}
\hypersetup{colorlinks=true, citecolor=blue, urlcolor=blue, linkcolor=blue}
\begin{document}
\title{Dynamical reciprocity in interacting games: numerical results and mechanism analysis}

\author{Rizhou Liang}
\affiliation{School of Physics and Information Technology, Shaanxi Normal University, Xi'an 710062, People's Republic of China}
\author{Qinqin Wang}
\affiliation{School of Physics and Information Technology, Shaanxi Normal University, Xi'an 710062, People's Republic of China}
\author{Jiqiang Zhang}
\affiliation{School of Physics and Electronic-Electrical Engineering, Ningxia University, Yinchuan 750021, People's Republic of China}
\affiliation{Beijing Advanced Innovation Center for Big Data and Brain Computing, Beihang University, Beijing 100191, People's Republic of China}
\author{Guozhong Zheng}
\affiliation{School of Physics and Information Technology, Shaanxi Normal University, Xi'an 710062, People's Republic of China}
\author{Lin Ma}
\affiliation{School of Physics and Information Technology, Shaanxi Normal University, Xi'an 710062, People's Republic of China}
\author{Li Chen}
\email[Email address: ]{chenl@snnu.edu.cn}
\affiliation{School of Physics and Information Technology, Shaanxi Normal University, Xi'an 710062, People's Republic of China}
\affiliation{Robert Koch-Institute, Nordufer 20, 13353 Berlin, Germany}

\date{\today }

\begin{abstract}
We study the evolution of two mutually interacting games with both pairwise games as well as the public goods game on different topologies. On 2d square lattices, we reveal that the game-game interaction can promote the cooperation prevalence in all cases, and the cooperation-defection phase transitions even become absent and fairly high cooperation is expected when the interaction goes to be very strong. 
A mean-field theory is developed that points out new dynamical routes arising therein. Detailed analysis shows indeed that there are rich categories of interactions in either individual or bulk scenario: invasion, neutral, and catalyzed types; their combination puts cooperators at a persistent advantage position, which boosts the cooperation. The robustness of the revealed reciprocity is strengthened by the studies of model variants, including asymmetrical or time-varying interactions, games of different types, games with time-scale separation, different updating rules etc. The structural complexities of the underlying population, such as Newman--Watts small world networks, Erd\H{o}s--R\'enyi random networks, and Barab\'asi--Albert networks, also do not alter the working of the dynamical reciprocity. In particular, as the number of games engaged increases, the cooperation level continuously improves in general. Our work thus uncovers a new class of cooperation mechanism and indicates the great potential for human cooperation where concurrent issues are so often seen in the real world. 
\end{abstract}
\pacs{87.23.Ge, 02.50.Le, 89.75.Fb, 87.23.Kg}
\maketitle

\section{Introduction}\label{sec:introduction}
The rise of human civilization is built upon the widespread cooperation at almost every corner of socioeconomical and other activities~\cite{axelrod2006evolution}. Its decay, by contrast, generally leads to regress of human welfare or even wars. The recent decades have witnessed some imminent crises such as global warming, trade wars, and more recently COVID-19~\cite{Covid19} etc. Solutions to any of them require multilateral cooperation, a throughout understanding of what motivates cooperation and how it evolves, and why it fails, is then needed. According to the Darwinism, however, natural selection favors the fittest, those who are altruistic incur a cost to themselves, leading to a less chance to survive. Cooperation is not a reasonable option by logic. Revealing the hidden mechanisms behind is therefore of fundamental importance and has been listed as one of the grand scientific challenges within this century~\cite{Pennisi2005How}.  

Within the framework of evolutionary game theory, many important progresses have been made~\cite{Nowak2006Five,Perc2017Statistical}. By analyzing those canonical models such as prisoner's dilemma (PD), the snowdrift game (SG), the public good game (PGG), and the collective-risk dilemma etc, valuable insights are obtained and several mechanisms are revealed. These include direct~\cite{trivers1971evolution} and indirect reciprocity~\cite{nowak1998evolution}, kin~\cite{hamilton1964genetical} and group selection \cite{keller1999levels,Queller1964Group}, spatial or network reciprocity \cite{nowak1992evolutionary}, reward and punishment \cite{Sigmund2001Reward}, social diversity and hierarchy~\cite{Santos2005Scale-Free,Santos2008Social,Liang2021Social}, and so on. In particular, it is found that the presence of population structures, either static or dynamic, is able to promote cooperation compared to the well-mixed scenario, because cooperators form clusters that can prevail against the invasion of defectors. This so-called network reciprocity has been extensively studied in the past few decades~\cite{szabo2007evolutionary,Wang2015Evolutionary}.

Alongside these theoretic insights, recent behavioral experiments also expand our understanding of cooperation~\cite{Rand2013Human}. As a new paradigm, the recruiting volunteers are configured with some given topologies and rules, they are well-motivated to play the games. But due to the complexities like human psychology, cultures or personalities etc, inconsistencies with theoretic predictions are often unveiled~\cite{Dirk2012Conditional}. For example, experiments with static structured population do not found their advantage in promoting cooperation in general as the network reciprocity predicts~\cite{traulsen2010human,Carlos2012Heterogeneous,Gracial2012Human}. Some extra conditions regarding the game experiment settings have to be satisfied for cooperation to thrive~\cite{Rand2014Static}. These facts imply that some essential factors could be missing in the most of current game-theoretic models, and the experiment-driven model improvement effort is required in the future.  

As a common practice, most of the existing works mainly focus on a single game scenario, with a belief in the mind that when the dynamics of a single game is well-understood, the wisdom obtained is supposed to be applicable to many other situations, even with more games being evolved. The philosophy behind is the reductionism essentially. In the real world, however, entities are generally evolved in multiple games simultaneously, where they could influence each other's evolution. For example, colleagues potentially work in a couple of concurrent projects, nations are involved in several issues such as trade, security, culture etc. In these contexts, the progress or conflicts in one issue is likely to affect the evolution of others, they have a stake in each other explicitly or implicitly. In non-human species, field research have made similar observations. For example, in chimpanzee societies, their activities such as grooming, hunting, sharing meat, supporting one another in conflicts, border patrols etc are found to be closely correlated with each other, e.g. a male chimpanzee with good hunting skill is more likely to be groomed by others and vice versa~\cite{hammerstein2003genetic}.

Till now, only a few models have considered the multiple games, but with different emphases. One line is along the multi-game dynamics. An early conceptually related work is the study by Cressman et al. \cite{cressman2000evolutionary} where two 2-strategy games are played and the eventual states can be described by the dynamics of the separate game. 
 Later research show that the fate of a single game generally cannot be determined without incorporating the messages of other games \cite{chamberland2000example, hashimoto2006unpredictability,venkateswaran2019evolutionary}, e.g. persistent cycles could arise within coupled one-equilibrium games. These works are done in a mean-field sense, and only consider one-shot game scenario where the potential reciprocity in evolution is beyond their scope.  
 
Another line is within the framework of interdependent networks, where different games are played on different layers of networks and they are coupled by means of the payoff/utility function~\cite{wang2013interdependent,gomez2012evolution,wang2012probabilistic,wang2013optimal,santos2014biased, wang2014evolutionary,battiston2017determinants,wang2014degree,jin2014spontaneous}.
 Zhen et al. \cite{wang2013interdependent} consider two PGGs being played on two symmetrically connected lattices, and the utility function includes not only the contribution of the payoff of the focal site plus its neighborhood's payoffs, but also the contributions of the neighborhood in the other lattice. They found that as the neighborhoods' contribution in both lattices increases, the cooperation level is promoted.  
 G\'omez-Gardenes et al. \cite{gomez2012evolution} extend the study to arbitrary number of layers where they adopt Erd\H{o}s-R\'eny random graphs and PD for each layer, and the net payoff is through the equal contribution among all layers used for the strategy updating. Their work shows a resilience of cooperation for extremely large values of temptation to defect and this resilience is intrinsically related to a non-trivial organization of cooperation across different layers. 
Asymmetrical game settings are also studied \cite{santos2014biased, wang2014evolutionary}, where different games are unfolded on different layers. For example, Santos et al. \cite{santos2014biased} considers PD and SG being posed on two layers of regular random networks respectively, individuals imitate neighbors from the same layer with a probability, and neighbors from the second layer with a complementary probability. Therefore the strategy transfer is allowed between layers and they find that while such coupling is able to promote the cooperation in the PD layer, but it is detrimental for the cooperation in the SG layer. 
Within the framework of interdependent networks, its specific construction matters~\cite{wang2012probabilistic,wang2013optimal,battiston2017determinants,wang2014degree}. In particular, the link overlap across different layers is shown to be crucial~\cite{battiston2017determinants}, there is no benefit for cooperation if without any structural correlation. The impact of other topological effects such as degree mixing is also studied in \cite{wang2014degree}, and some other dynamical processes like spontaneous symmetry breaking between different layers are uncovered in \cite{jin2014spontaneous}. 
All these promotions are attribute to the \emph{interdependent network reciprocity}~\cite{wang2013interdependent}, a subcategory of network reciprocity. Intuitively, one can view it as an efficient construction of population relationship that tends to maximize the previously uncovered network reciprocity. 

The most related work is by Donahue et. al~\cite{donahue2020evolving}, where they propose a framework termed multichannel games. Each channel represents a repeated game and players interact over multiple channels and these channels are dependent with each other. Based on two donation games, they uncovered the evolutionary advantage of cooperation due to the game linkage. This finding acts as a good starting point towards a new fundamental category of reciprocity --- \emph{dynamical reciprocity}, as a counterpart of the well-studied network reciprocity. It emphasizes that the reciprocity stems from the game-game dynamics rather than the underlying topology of population. A bunch of fundamental questions, however, remain unanswered: \emph{what's the typical evolutionary dynamics when more games are engaged? how robust is the dynamical reciprocity? what's mechanism behind this new type of reciprocity?} These are what we are trying to answer in this work.

The aim of the present paper is to introduce and formalize the interacting games and systematically investigate the impact of game interplay on the cooperation prevalence. Different games interact through a perceived payoff, a function of payoffs in all games. To our surprise, we are not only able to see the cooperation promotion, but also the cooperation-defection phase transitions could disappear, where an absorbing state of nearly full cooperation is approached. More numerical studies show this sort of promotion is quite robust, confirmed by variants with both different game dynamics (the types of game, the updating rule and synchronicity, the game coupling etc.) and different underlying topologies. A mean-field treatment indicates that new dynamical routes towards cooperation come into play, which is confirmed by detailed mechanism analysis. There, apart from the invasion category of interaction that is also present in the single game case, two other new categories of interactions --- neutral and catalyzed --- are identified. Working together, these three categories of interactions lead to a persistent advantage of cooperators over the defectors. The promotion is further enhanced when more games are engaged, where the revealed mechanism still holds.

The paper is organized as follows:
In Sec.~\ref{sec:model}, we formulate the interacting games of arbitrary number. 
In Sec.~\ref{sec:results}, the preliminary results of two symmetrically interacting games on 2d regular lattice are shown.  
In Sec.~\ref{sec:MF}, we present a mean-field treatment within the framework of replicator equations to see what dynamics would arise in the presence of game-game interactions.
In Sec.~\ref{sec:mechanism}, the dynamical mechanism of the revealed reciprocity is discussed in details by classifying all interacting pairs.
More numerical results regarding the robustness of the revealed reciprocity are provided regarding dynamical variations in Sec.~\ref{sec:robustness}, 
and structural variations in Sec.~\ref{sec:networks}. More games are considered in Sec.~\ref{sec:more} to extrapolate what would be expected when the number of games increases.
Finally, some concluding remarks are given in Sec.~\ref{sec:summary}, together with the implications and some open questions being listed.

A short version of the work is presented in a short letter~\cite{CSL}.

\begin{figure}
\centering
\includegraphics[width=0.9\linewidth]{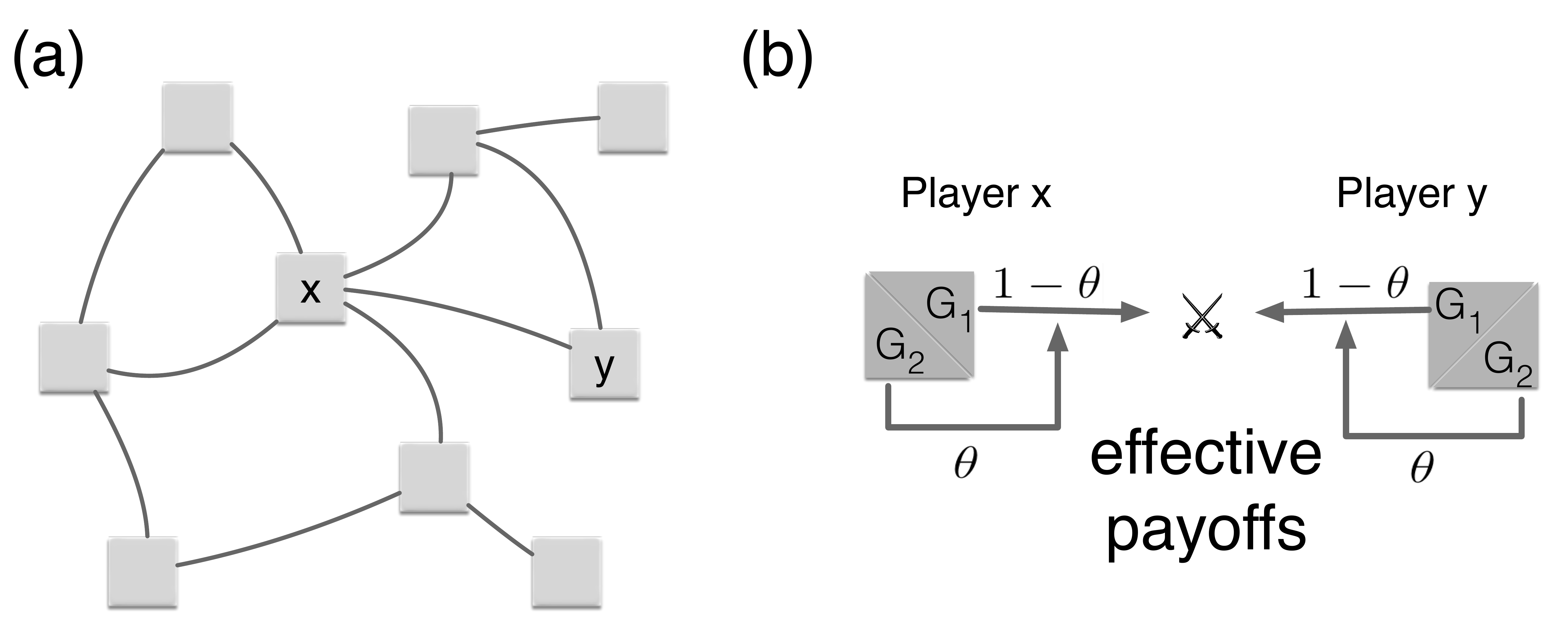}
\caption{
Modeling two interacting games.
(a) Consider a group of networked players, they play two games $G_{1,2}$ simultaneously, two payoffs are obtained accordingly, which can be interpreted as the fitness in their evolution.
(b) When two neighboring individuals, say player $x$ and $y$, are to update their strategies with respect to a given game (e.g.  game $G_1$ here), the update not only depends on the payoffs obtained in $G_1$ (with weight $1-\theta$), but also the one in the other game $G_2$ (with $\theta$), this combination is termed as the effective payoffs, see Eq.(\ref{eq:effective_linear}).
}
\label{fig:scheme}
\end{figure}

\section{General formulation}\label{sec:model}
We formulate the interacting games within the evolutionary framework for where $m$ games are denoted as $\mathbb{G}=\{G_1,G_2,...,G_m\}$, they are played simultaneously in the population of size $N$ sharing the same underlying topology. The corresponding strategy set is denoted as $\mathbb{S}_m$.  Assume that each player could be in one of two states in each game: cooperation (C) or defection (D), i.e. $\mathbb{S}_1=\{C,D\}$. By combination there are $2^{m}$ elements in $\mathbb{S}_m$, e.g. the strategy set for $m=2$ is $\mathbb{S}_2=\{XY|CC, CD, DC, DD\}$, where $X$, $Y$ correspond to the state in game $G_1$, $G_2$, respectively.

In our study, we adopt the general pairwise games (GPG) and public goods game (PGG). 
The GPG is defined as follows: when both players cooperate, each gets a reward $R$; when both defect, then each gets a punishment $P$; and the mixed encounter yields a temptation $T$ for the defector while the cooperator becomes a sucker with a payoff $S$. Different rankings of the four payoffs lead to different game types. Specifically, four types of games are defined in $S-T$ parameter space by fixing $R\!=\!1$ and $P\!=\!0$: i) $1<T<2$ and $0<S<1$ for the snowdrift game (SG); ii) $0<T<1$ and $0<S<1$ for the harmony game (HG); iii) $0<T<1$ and $-1<S<0$ for stag hunt game (SH); iv) $1<T<2$ and $-1<S<0$ for prisoner's dilemma (PD). The PGG can be considered as an extension of PD where an arbitrary number of players can play together in a group. It is defined as follows: in each round, every player in the group chooses either to contribute 1 to the common pool as a cooperator, or nothing as a defector; the sum of the contribution is then multiplied by a gain factor $r>1$, reflecting the synergetic effect; finally, the resulting amount of benefit is equally shared among all members in the group, including those defectors. Therefore the net payoff of cooperators have to subtract 1 from the shared amount, whereas the defectors don't need to do so. If without any mechanism, defection is preferred.

The system is initialized with random conditions if not stated otherwise where each player randomly chooses to cooperate or defect in each game. The evolution follows the standard Monte Carlo (MC) procedure. At an elementary step, a random player $x$ is chosen and a random game $g\in \mathbb{G}$ is played, then we compute its payoff $\Pi_x$ according to the game setting. Next, one of its neighbors $y\in\Omega_x$ is picked randomly, also its payoff $\Pi_y$ is computed. Lastly,  player $x$ adopts $y$'s strategy according to some function $W(s^g_y\rightarrow s^g_x)=W(\Pi_x,\Pi_y)$ that translates their payoff difference into the learning propensity. In our study, the Fermi rule~\cite{szabo1998evolutionary} is adopted as 
\begin{equation}
W(s_y^g\rightarrow s_x^g)=\frac{1}{1+\exp[(\widehat{\Pi}_x^g-\widehat{\Pi}_y^g)/K]} ,
\label{eq:fermi}
\end{equation}
where $K$ is a temperature-like parameter, measuring the uncertainties in the strategy adoption, its inverse can be interpreted as the selection pressure. $K$ is fixed at 0.1 throughout the work if not stated otherwise. $\widehat{\Pi}_x^g$ is the \emph{effective payoff} perceived by player $x$ in game $g$ defined below that is used to update its strategy. A full MC step is comprised of $m\times N$ such elementary steps, meaning that every player is going to update its strategy once for each game on average.

The effective payoff $\Pi^g_x$ that is formally defined as
\begin{equation}
\widehat{\Pi}_x^g=E({\Pi}_x^{G_1},{\Pi}_x^{G_2},...{\Pi}_x^{G_m};P^g(\theta_1,\theta_2,...,\theta_m)),
\label{eq:effective_m}
\end{equation}
by which the game-game interactions come into play. It captures the facts that the decision-making of a given game $g$ would base upon a perceived payoff through integrating payoffs in all games instead of simply the one under play. Here $\theta_i\in[0,1]$ is the contribution weight of game $G_i$.
The distribution $P^g(\theta_1,\theta_2,...,\theta_m)$ then determines to how $m$ games influence the perceived payoffs in game $g$. 

Note that, the implemented MC simulation procedure is to mimic the continuous-time evolution as in the real world, where the strategy updating is asynchronous. To our interest, we also consider synchronous updating (SU) scheme as follows. All $m$ games are repeatedly played in circular order $G_1,G_2,...,G_m,G_1,...$; for a given game, all players simultaneously compute and compare their payoffs to one randomly chosen neighbor, and make strategy update according to Eq.(\ref{eq:fermi}). $m$ such discrete steps in the SU scheme guarantee that every game is precisely played once for each player. 

Four more different updating rules~\cite{roca2009evolutionary}: Moran-like rule, replicator rule, multiple replicator rule, and the unconditional rule are also investigated in Sec.~\ref{sec:robustness} for robustness studies. 

We adopt 2d square lattices with the size of $N=L\times L$ as the underlying structure for most studies, where each individual plays the games with its four nearest neighbors, and the periodic boundary condition is assumed. We will also study Newman-Watts networks, Erd\H{o}s-R\'enyi networks, and Barab\'asi--Albert networks in Sec.~\ref{sec:networks}. As noted above that all games are assumed to share the same set of links, this is not very realistic of course, since many games are unfolded on their own sets of connections, as in the multiplex networks. However, these structural intricacies in multiplex networks would bring additional complexities that will confound our understanding of the reciprocity purely from the dynamical part, therefore we would like to avoid them.

\begin{figure}
\centering
\includegraphics[width=0.8\linewidth]{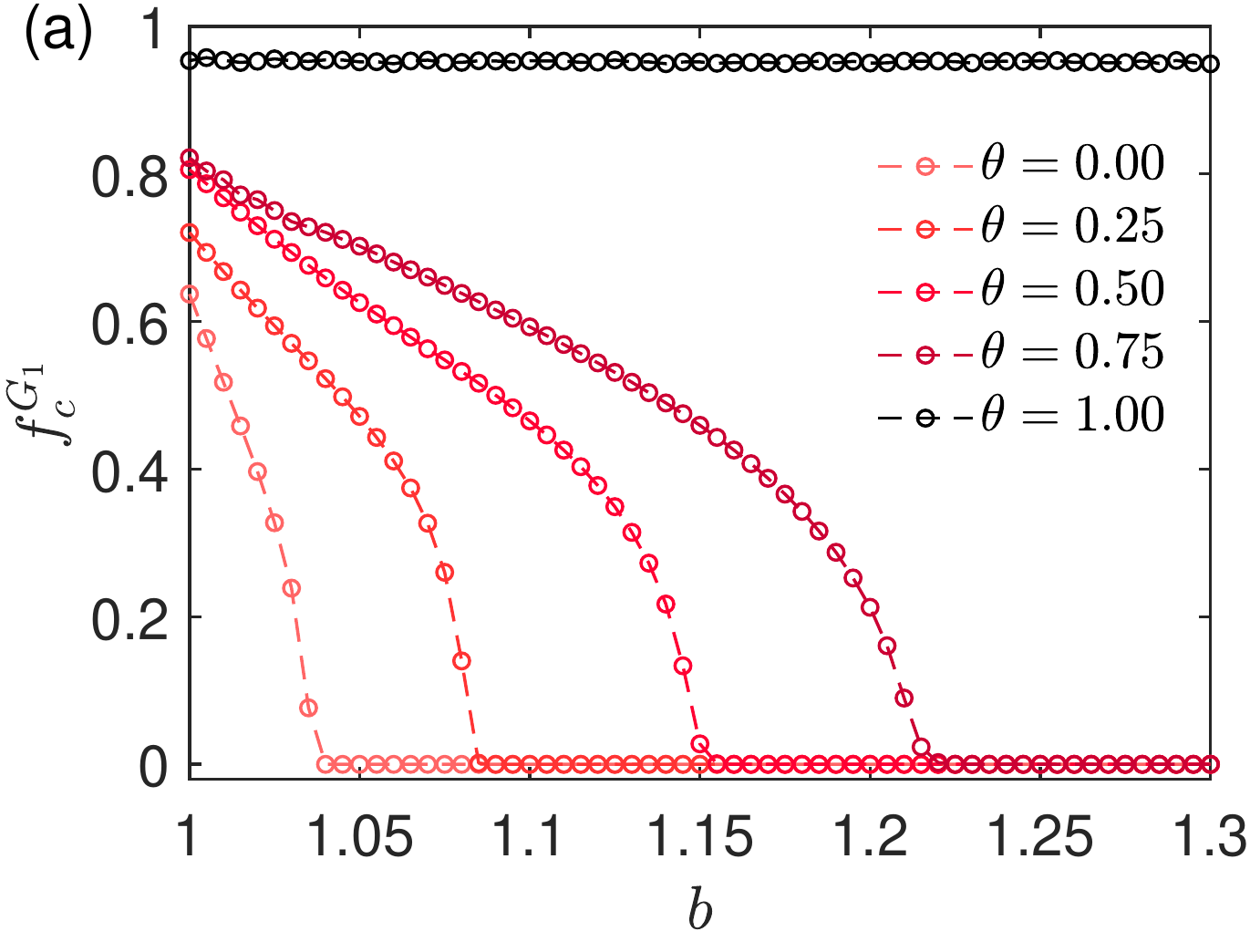}
\includegraphics[width=0.8\linewidth]{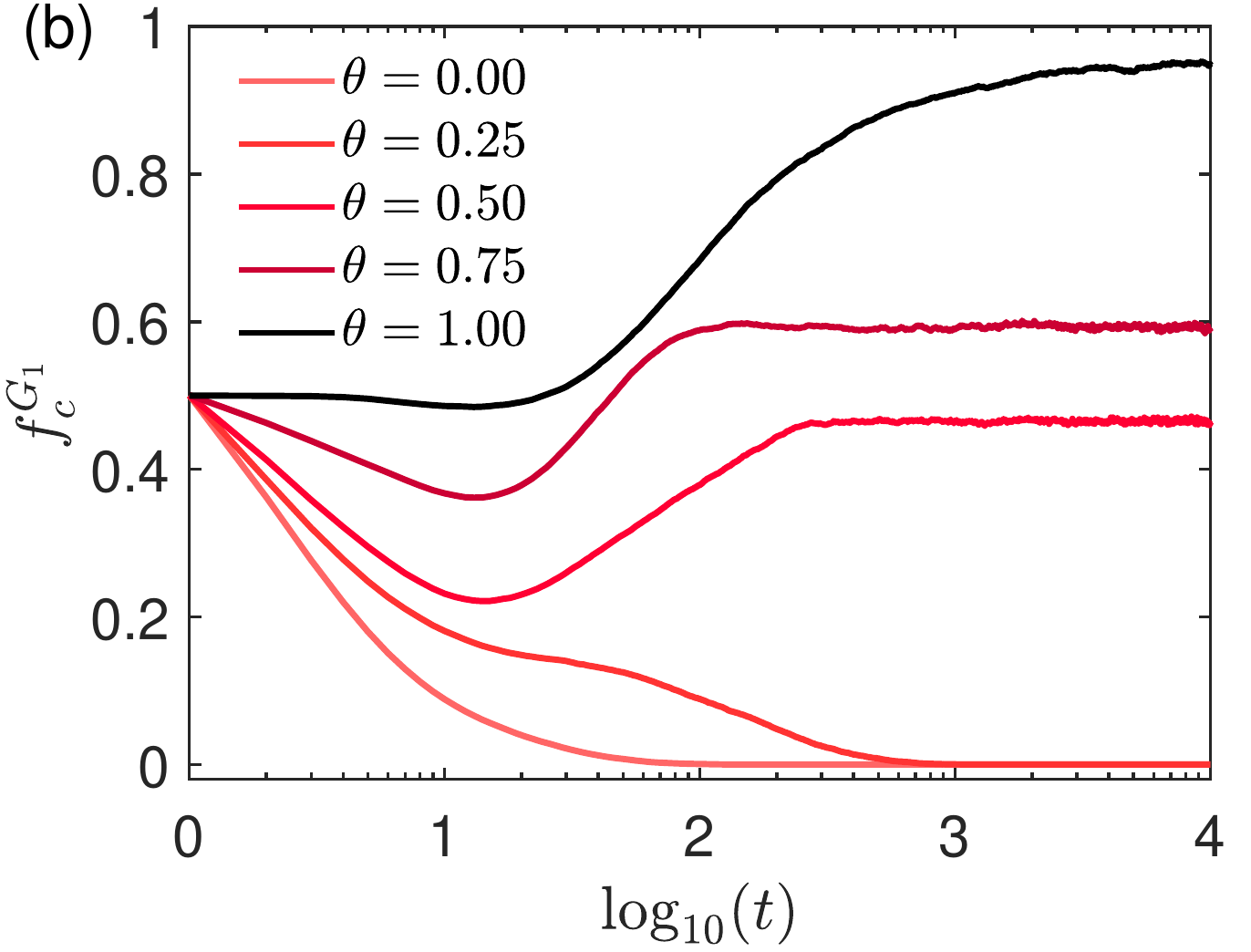}
\caption{
The evolution of two symmetrically interacting PD on the 2d square lattice.
(a) The phase transitions of cooperation prevalence regarding game $G_1$ ($f^{G_1}_C=f_{CC}+f_{CD}$) for different interaction strengths $\theta$, the temptation $b$ is the control parameter.  
(b) Time series for $b=1.1$. 
Note that, $f^{G_2}_C\approx f^{G_1}_C$ due to the symmetry and is not shown.
Parameters: $L=1024$ and $K=0.1$, the random initial condition for both games, each point is averaged over 50 ensembles after transient in (a).
}
\label{fig:ts}
\end{figure}

\begin{figure*}[htbp]
\centering
\includegraphics[width=0.8\linewidth]{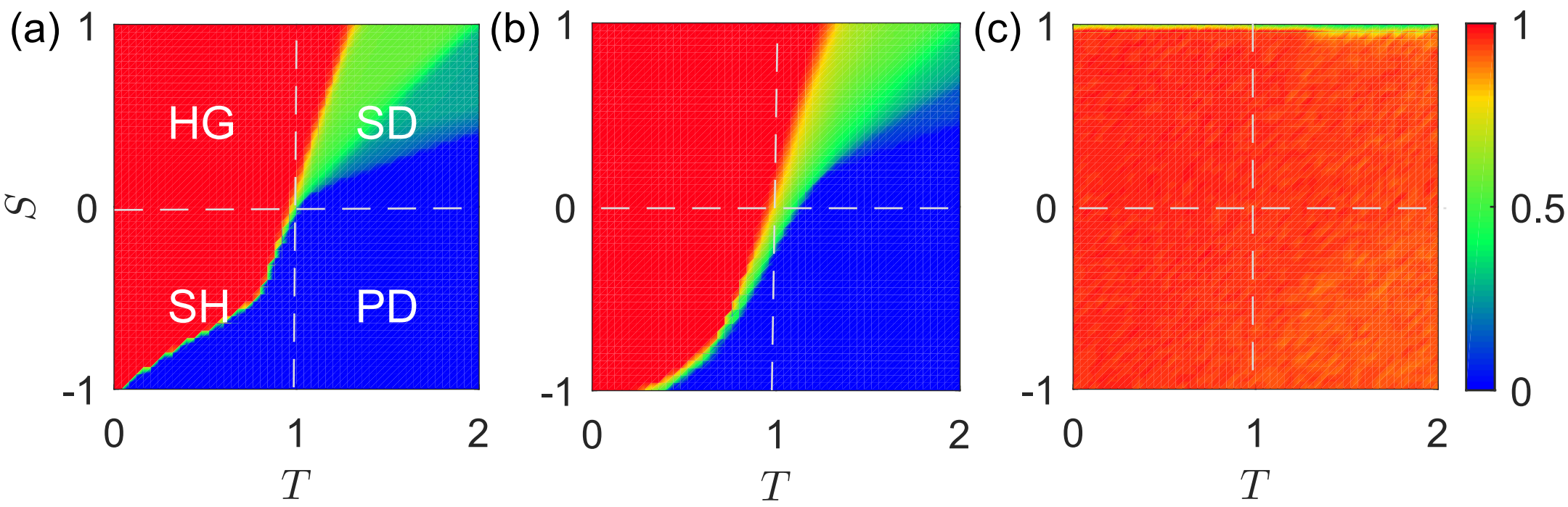}
\caption{(Color online)
Color-coded fraction of cooperators regarding the first game ($f^{G_1}_C=f_{CC}+f_{CD}$) for the general pairwise game within the $S-T$ parameter space with $\theta=0$, 0.5, and 1 on the 2d square lattice, respectively shown in (a-c). Due to the symmetry, $f^{G_2}_C\approx f^{G_1}_C$. Four quadrants correspond to four different games (defined in Sec.~\ref{sec:model}).
Parameters: $R=1$, $P=0$, $L=128$.  
}
\label{fig:GPG}
\end{figure*}
\section{Preliminary results for 2 interacting games on 2d square lattice}\label{sec:results}
In this section, we shall only discuss the case of two interacting games $\mathbb{G}=\{G_1,G_2\}$ with perfect symmetry, they are unfolded on a 2d square lattice (see Fig.~\ref{fig:scheme}).  A simple linear combination is used for the effective payoff as
\begin{equation}
\widehat{\Pi}_{x}^{G_{1,2}}=(1-\theta)\Pi_{x}^{G_{1,2}}+\theta \Pi_{x}^{G_{2,1}},
\label{eq:effective_linear}
\end{equation}
where $E(...)$ is a linear function, $P^{G_1}(\theta_1,\theta_2)=\{1-\theta,\theta\}$ and $P^{G_2}(\theta_1,\theta_2)=\{\theta,1-\theta\}$ are also symmetrical in Eq.~(\ref{eq:effective_m}). Here we interpret the contribution weight $\theta$ as the game-game \emph{interaction strength}. The larger the value of $\theta$, the stronger impact of the other game is posed. The case of $\theta=0$ reduces the model into two independent games, while the other extreme $\theta=1$ corresponds to the cross-playing scenario where the decision-making of a given game is entirely determined by the payoffs in the other game. More often cases in reality are supposed to occur in between. These constitute a parsimonious model of two interacting games.

A typical example of GPG is  two symmetrically interacting PD games reported in~\cite{CSL}, see Fig.~\ref{fig:ts}. 
A monotonic promotion of cooperation is observed as the interaction strength $\theta$ increases. As the case of cross-playing is adopted, the cooperation prevalence becomes independent of the game parameter $b$ any more. The time series support these observations, and in particularly show that the decay in the initial stage is also become lesser when $\theta$ become larger. This means that, even before clusters are formed, the game interaction provide some protections for cooperators from being exploited by defectors compared to the case of independent game case, where the cooperators are invaded and go to be extinct for the given parameter in Fig.~\ref{fig:ts}(b). 

We further show that this promotion is universal for all game types within the GPG formulation, see Fig.~\ref{fig:GPG}. We see that as $\theta$ increases, the defection region shrinks. In particular, when $\theta\rightarrow1$, the cooperation maintains at a fairly high level ($> 0.8$) for the whole parameter $S-T$ space, the difference among the four games nearly disappears. 

\begin{figure}[htbp]
\centering
\includegraphics[width=0.8\linewidth]{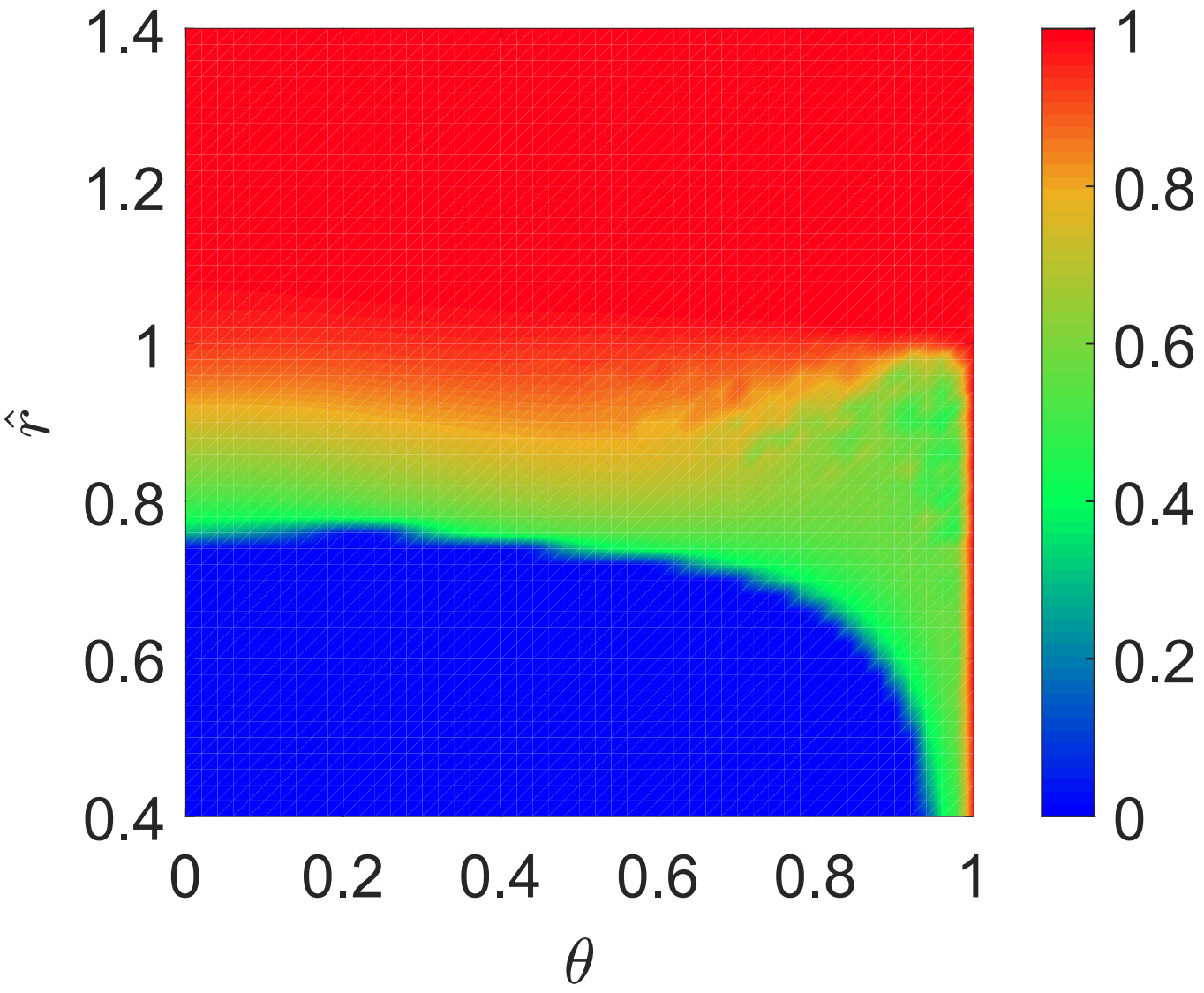}
\caption{(Color online)
Color-coded fraction of cooperation prevalence $f^{G_1}_C$ in $\theta-\hat{r}$ parameter space for two symmetrically interacting public goods games. $\hat{r}=r/(k+1)$ is the normalized gain factor, and $k+1$ is the number of games the individual is evolved.
Also $f^{G_2}_C\approx f^{G_1}_C$ due to the symmetry. Here $L=128$ and $K=0.5$.
}
\label{fig:PGG}
\end{figure}

In fact, a similar observation is also made in two symmetrically interacting PGGs.
Fig.~{\ref{fig:PGG}} shows the prevalence of cooperation within a two-parameter space. While a larger value of normalized gain factor $\hat{r}$ tends to raise the cooperation propensity, a stronger interaction strength $\theta$ generally facilitates cooperation as well.
Together with Fig.~\ref{fig:GPG}, we conclude that as the game-game interaction becomes stronger, cooperation continues to improve, and fairly high cooperation is seen as $\theta\rightarrow1$, irrespective of the game type.

\section{A mean-field analysis}\label{sec:MF}
In theory, the evolutionary games can be described by a mean-field treatment based on the replicator equation (RE) \cite{roca2009evolutionary}, which was introduced in 1978 by Taylor and Jonker \cite{taylor1978evolutionary}. RE characterizes the evolution of frequencies or fractions of different species in the population by taking into account their mutual influence on each other's fitness. Mathematically, it successfully captures the selection process and provides a bridge between the Nash equilibrium in static payoff matrix and the evolutionary stable strategies in evolution. 

However, there are some assumptions required explicitly or implicitly in the derivation of RE as follows. (1) The population is infinitely large; (2) The population is well-mixed so that each individual interacts with an equal probability with everyone else; (3) No mutation is allowed for the strategies, their frequency changes are only due to the reproduction; (4) The evolution of frequencies is linearly proportional to their fitness difference. Derivations from the finite-size effect or from the structured property in the real population are expected according to assumptions (1) and (2).

With these assumptions, let's consider a population with $m$ games, their evolution can formally be described as 
\begin{equation}
\dot{{f}_{s}}=f_{s}(\Pi_{s}-\bar{\Pi}),\label{eq:MF}
\end{equation}
where $f_{s}$ is the frequency or fraction of population within state $s\in\mathbb{S}_m$, $\Pi_{s}$ is its fitness $\Pi_{s}=\sum_{g\in\mathbb{G}}\Pi^{g}_s$, and $\bar{\Pi}=\sum_{s}f_{s}\Pi_{s}$ is the average fitness of the whole population. The species with a high fitness tends to increase its fraction, the one with a low value tends to reduce instead.

\subsection{A single pairwise game}\label{subsec:mf_single}
Let's first recall the well-known single pairwise game case, where $m\!=\!1$ and $\mathbb{S}_1\!=\!\{C,D\}$. $f_{C,D}$ are the fraction of cooperators
and defectors respectively, with $f_{C}+f_{D}=1$. By applying the payoff matrix given in Sec. II, 
the RE is then 
\begin{eqnarray}
\dot{{f}}_{C}& = & f_Cf_D[f_C(R-T)+f_D(S-P)] \nonumber \\
                   & = & f_{C}(1-f_{C})(\Pi_C-\Pi_D),\label{eq:MF_single}
\end{eqnarray}
where $\Pi_C=R f_C+S(1-f_C)$ and $\Pi_D=T f_C+P(1-f_C)$. There are three fixed points:
\begin{equation}
f_{C}^{*}=0,1,\frac{{P-S}}{R+P-T-S},\label{eq:MF_solution}
\end{equation}
which correspond to full defection, full cooperation, and a mixed strategy, respectively.
Their stabilities are determined by doing linearization of Eq. (\ref{eq:MF_single}) and computing the corresponding eigenvalues around these fixed points. Only negative eigenvalues guarantee a stable solution. Well-known facts of the stable solution are as follows~\cite{smith1982evolution}: full defection for PD, mixed strategy for SD, coexistence of full cooperation and full defection for SH depending on the initial condition, and full cooperation for HG. 

\subsection{Two interacting pairwise games}\label{subsec:MF2}
Now let's extend the RE treatment to the case of two symmetrically interacting pairwise games, where $\mathbb{S}_2=\{CC,CD,DC,DD\}$. The four fractions satisfy $\sum_s f_s=1, s\in\mathbb{S}_2$.
The overall fitness of a given state $s$ is $\Pi_{s}=\sum_{g}\Pi_{s}^{g}=\Pi_{s}^{G_1}+\Pi_{s}^{G_2}$. Along the setup in numerical simulations, the two games are symmetrical both in the game parameterization and interactions. Without game interactions, the fitness of the four species in the two games are
\begin{equation}
\begin{pmatrix}
\Pi^{G_1}_{CC} \\ \Pi^{G_1}_{CD} \\ \Pi^{G_1}_{DC} \\ \Pi^{G_1}_{DD} 
\end{pmatrix}
=
\begin{pmatrix}
R & R & S & S \\
R & R & S & S \\
T & T & P & P \\
T & T & P & P 
\end{pmatrix}
\begin{pmatrix}
f_{CC} \\ f_{CD} \\ f_{DC} \\ f_{DD} 
\end{pmatrix},
\label{eq:MF_f1}
\end{equation}
and 
\begin{equation}
\begin{pmatrix}
\Pi^{G_2}_{CC} \\ \Pi^{G_2}_{CD} \\ \Pi^{G_2}_{DC} \\ \Pi^{G_2}_{DD} 
\end{pmatrix}
=
\begin{pmatrix}
R & S & R & S \\
T & P & T & P \\
R & S & R & S \\
T & P & T & P 
\end{pmatrix}
\begin{pmatrix}
f_{CC} \\ f_{CD} \\ f_{DC} \\ f_{DD} 
\end{pmatrix}.
\label{eq:MF_f2}
\end{equation}

With these, we define the \emph{effective fitness} analogously as 
\begin{equation}
\begin{pmatrix}
\widehat\Pi^{G_1}_{s} \\ \widehat\Pi^{G_2}_{s} 
\end{pmatrix}
=
\begin{pmatrix}
1\!-\!\theta & \theta \\
\theta & 1\!-\!\theta\!  
\end{pmatrix}
\begin{pmatrix}
\Pi^{G_1}_{s} \\ \Pi^{G_2}_{s} 
\end{pmatrix},
\label{eq:effective_fitness}
\end{equation}
corresponding to the effective payoffs as in Eq.  (\ref{eq:effective_linear}) in the numerical simulations.
The RE Eq. (\ref{eq:MF}) is then rewritten as 
\begin{equation}
\dot{{f}}_{s}=f_{s}(\widehat{\Pi}_{s}-\bar{\Pi}),
\label{eq:MF_RE}
\end{equation}
where $\widehat{\Pi}_{s}=\sum_{g}\widehat{\Pi}_{s}^{g}=\widehat{\Pi}_{s}^{G_1}+\widehat{\Pi}_{s}^{G_2}$. And the mean fitness is
$\bar{\Pi}=\sum_{s}f_{s}\widehat{\Pi}_{s}$.

With some algebra (see Appendix~\ref{app:A} for details), we can derive the evolution of cooperator fraction with regard to either game, say game $G_1$ ($f^{G_1}_C=f_{CC}+f_{CD}$), as follows 
\small
\begin{equation}
\dot{f}^{G_1}_{C}=f^{G_1}_{C}f^{G_1}_{D}(\Pi^{G_1}_C-\Pi^{G_1}_D) + (f_{CC}f_{DD}-f_{CD}f_{DC})(\Pi^{G_2}_C-\Pi^{G_2}_D), 
\label{eq:mf_cor}
\end{equation}
\normalsize
where $\Pi^{G_1}_C\!=\!\Pi^{G_1}_{CC}\!=\!\Pi^{G_1}_{CD}$, $\Pi^{G_1}_D\!=\!\Pi^{G_1}_{DD}=\Pi^{G_1}_{DC}$, $\Pi^{G_2}_C=\Pi^{G_2}_{CC}=\Pi^{G_2}_{DC}$, and $\Pi^{G_2}_D=\Pi^{G_2}_{DD}=\Pi^{G_2}_{CD}$. The first term of the rhs. is exactly the same as in the single game dynamics shown in Eq. (\ref{eq:MF_single}). It means that the cooperator fraction in game $G_1$ tends to increase when $\Pi^{G_1}_C>\Pi^{G_1}_D$.
The new dynamics lies in the second term, which captures the game-game interaction via the interaction pairs of CC -- DD and CD -- DC.  Specifically, when CC individuals meet up with DD ones, if their fitness satisfies $\Pi^{G_2}_C>\Pi^{G_2}_D$ the advantage of cooperators in game $G_2$ transfer DD to be CD due to the game correlation. By contrast, when CD individuals meet up with DC ones, this advantage is instead to reduce the cooperation prevalence, transferring CD to DD. But for the way around (i.e. $\Pi^{G_2}_D>\Pi^{G_2}_C$), however, the advantage of defectors in game $G_2$ leads to the growth of cooperators in $G_1$, transferring DC to CC, which is unexpected when games evolve independently. Therefore, the above equation explicitly captures the cooperation dynamics from both intra- and inter-game interactions. Note that, by the exchange of game label 1 and 2 in Eq. (\ref{eq:mf_cor}), the equation describes the cooperator fraction evolution of game $G_2$, i.e. $f^{G_2}_{C} = f_{CC}+f_{DC}$.

\begin{figure}[b]
\centering
\includegraphics[width=0.7\linewidth]{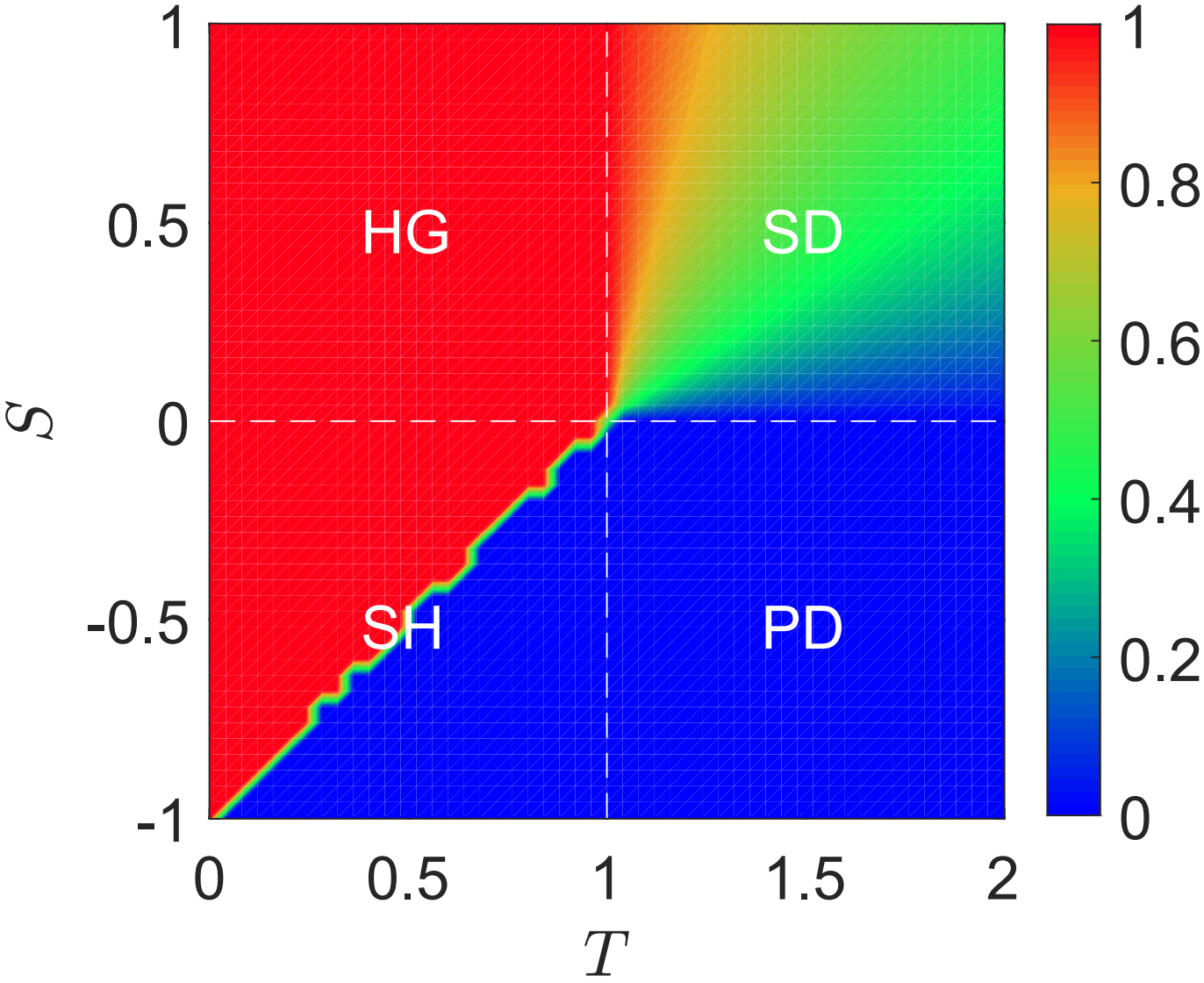}
\caption{(Color online)
The cooperator fraction $f^{G_1}_C$ for the two interacting pairwise games in the well-mixed population within $S-T$ parameter space. The resulting cooperation prevalence exactly corresponds to the solution of single game; the less significant bistability in SH game is simply due to the random initial conditions, where $f_C\approx f_D\approx 1/2$. 
Due to the symmetry, $f^{G_1}_C\approx f^{G_2}_C$ in all cases, except very few case along the line $S=T-1$ in SH, depending on their initial conditions. Other parameters: $R=1$, $P=0$, $N=2^{14}$. 
}
\label{fig:wellmixed}
\end{figure}

\begin{table*}[]
\begin{tabular}{@{}rlll@{}}
\hline\hline
\toprule
\multicolumn{1}{l}{} & \multicolumn{1}{c}{~~~~~~~~~~Individual interaction~~~~~~~~~~}                                                              & \multicolumn{1}{c}{~~~~~~~~~~~~~Bulk interaction~~~~~~~~~~~~~}                                                                    &  \\ \midrule
\hline
Invasion type~~        & \cellcolor[HTML]{EFEFEF}CC + DD $\xrightarrow{  G_1/G_2  }$ DC/CD + DD                                                                             
                                 & \cellcolor[HTML]{EFEFEF}CC + DD $\xrightarrow { G_1/G_2 }$ CC + CD/DC\\                                                                        &  \\
Neutral type~~         & \cellcolor[HTML]{EFEFEF}\begin{tabular}[c]{@{}l@{}}CC + DC $\xrightarrow{    G_1    }$ 2CC or 2DC\\ CC + CD $\xrightarrow{    G_2    }$ 2CC or 2CD \\ DD + CD $\xrightarrow{    G_1    }$ 2DD or 2CD\\ DD + DC $\xrightarrow{    G_2    }$ 2DD or 2CD\end{tabular} 
                               & \cellcolor[HTML]{EFEFEF}\begin{tabular}[c]{@{}l@{}}CC + DC $\xrightarrow{    G_1    }$ 2CC or 2DC\\ CC + CD $\xrightarrow{    G_2    }$ 2CC or 2CD\\ DD + CD $\xrightarrow{    G_1    }$ 2DD or 2CD\\ DD + DC $\xrightarrow{    G_2    }$ 2DD or 2CD   \end{tabular} &  \\\\
Catalyzed type~~       & \cellcolor[HTML]{EFEFEF}CD + DC $\xrightarrow{   G_1/G_2   }$ CD/DC + CC                                
                                   & \cellcolor[HTML]{EFEFEF}CD + DC $\xrightarrow{   G_1/G_2   }$ DC/CD + DD                                                                         
                                   &  \\ \bottomrule
\hline
\end{tabular}
\caption{Categories of interactions in 2 interacting PD games within the cross-playing scheme ($\theta=1$). Six pairwise interactions are classified into three categories in either individual or bulk scenarios.}
\label{tab:2games}
\end{table*}

By analytically solving the mean field Eqs. (\ref{eq:MF_RE}) (see Appendix~\ref{app:B}), we found that the stable solutions are exactly the same as the single game case, meaning the game-game interaction brings no impact on the cooperation evolution at the mean-field level. The two games are decoupled essentially in the well-mixed scenario. This means that the dynamical reciprocity from the game-game interaction should go hand in hand with network reciprocity --- it only works in the structured population.
Numerical simulations of two interacting PD games in a fully connected population confirm this conclusion as shown in Fig.~\ref{fig:wellmixed}, where by starting from random initial conditions, mixed strategies, full cooperation, coexistence of full cooperation and full defection, full defection are respectively seen in the quadrant I to IV, in accordance with the analytical results. In the following section, however, we show that the second term in Eq. (\ref{eq:mf_cor}) does play a role in the structured population, which brings fundamentally different dynamical routes towards cooperation.

\section{Dynamical mechanism}\label{sec:mechanism}
\subsection{The category of interactions for two games}\label{subsec:classification}
To understand the physics behind Eq. (\ref{eq:mf_cor}), and the dynamical reciprocity better, let's list all possible interactions in the two interacting games on 2d square lattices and classify them into different categories according to the net effect in their offspring reproduction (see Table \ref{tab:2games}). For simplicity, we focus on the cross-playing case ($\theta=1$) in pairwise games, where the classification is most clear, but qualitatively the following analysis can also be applied to the cases with $\theta<1$. 

Before proceed, we need to distinguish two interacting scenarios --- \emph{individual} and \emph{bulk} interactions. Consider two neighboring players $x$ and $y$, and assume $s_x\neq s_y$. In the individual scenario, we only compute the payoffs of the two players through the interacting pair $x-y$. This scenario applies to the context when their surroundings are unknown, like the random initial condition case, where the information of their neighbors is stochastic. Instead, when players of the same state are well-bulked, their payoffs can be explicitly estimated by incorporating both the intra- and inter-bulk gaming. Specifically for player $x$, its payoff is assumed to include the payoffs from its three neighbors with $s=s_x$ (intra-bulk) and the one from $x-y$ gaming (inter-bulk). These two scenarios are two extreme cases occurring in structured population, the actual evolution should occur somewhere in between. 

Since there are four different types of individuals, there are six interacting pairs by combination, which can be classified into the following three categories for either individual or the bulk interactions, see Table \ref{tab:2games}. 

\emph{(i) Invasion type} --- for the individual interaction, DD dominates over CC for either game thus the CC individual tends to become CD or DC; while when bulk CCs meet up bulk DDs, CCs at the interface have higher payoff than DDs (typically $3R>T$ for the lined up interface), CC bulks are supposed to invade into DD regions. In either case, the two cannot coexist and invasion happens, where the one at the disadvantageous position will be invaded and become CD or DC.

\emph{(ii) Neutral type} --- for both scenarios, there is a neutral outcome, where statistically no net change in cooperation is expected. Consider CC--CD interface as an example, the two individuals are of identical state regarding game $G_1$, $\Pi_{CC}^{G_1}\!=\!\Pi_{CD}^{G_1}\!=\!R$ and $4R$ for individual and bulk interactions, respectively. And since the evolution in game $G_2$ is determined by the difference in $\widehat{\Pi}_{CC,CD}^{G_2}$($= \Pi_{CC,CD}^{G_1}$), $\widehat{\Pi}_{CC}^{G_2}-\widehat{\Pi}_{CD}^{G_2}=0$ means that the evolution of CC--CD interface is neutral $W(CD\rightarrow DC)\!=\!W(DC\rightarrow CD)\!=\!1/2$ according to Eq.~(\ref{eq:fermi}), a random-walk-like movement of their boundaries is expected. This argument is applicable to the other three interfaces alike.

\emph{(iii) Catalyzed type} --- when a CD meets up a DC and they play game $G_1$; accordingly $\widehat{\Pi}_{CD}^{G_1}\!=\!\Pi_{CD}^{G_2}=T$ and $\widehat{\Pi}_{DC}^{G_1}\!=\!\Pi_{DC}^{G_2}=S$ with $T > S$, the DC individual is then likely to become CC by learning (i.e. $CD+DC \xrightarrow{\text{G}_1} CD+CC$). Analogously, when they play game $G_2$, the CD individual tends to become CC (i.e. $CD+DC \xrightarrow{\text{G}_2} CC+DC$). As a consequence, CC individuals are continuously produced. 
In the bulk scenario, however, the situation is reversed. 
When bulk CDs and bulk DCs meet up, at the lined-up interface $\widehat{\Pi}_{CD}^{G_1}= \Pi_{CD}^{G_2}=T$ and $\widehat{\Pi}_{DC}^{G_1}= \Pi_{DC}^{G_2}=3R$ with $T < 3R$, CD individuals then tend to become DD while DC remains unchanged when playing game $G_1$ (i.e. $CD+DC \xrightarrow{\text{G}_1} DD+DC$). And DC individuals likewise are likely to become DD when playing game $G_2$ (i.e. $CD+DC \xrightarrow{\text{G}_2} CD+DD$). Therefore, DDs are continuously produced instead for bulk interactions. 
In all these cases, the advantage or disadvantage regarding a given game, is not directly translated to any strategy update in the game \emph{per se}, but the state changes in the other game. This is reminiscent of catalyzed reactions in Chemistry. The presence of one game is to ``catalyze'' the evolution of the other game. They mutually catalyze each other' evolution. 

With this classification, it's straightforward to understand the impact of game-game interactions, or more specifically the second term in mean-field equation Eq. (\ref{eq:mf_cor}). First, since the four neutral interactions give rise to null net effect, no related term is included. Second, the CC--DD and CD--DC interacting pairs do have net effect in the cooperation evolution therefore they appear in the equation, but because they always have the opposite impact on cooperation in either individual or bulk scenario, they have opposite signs. Third, the mean-field treatment assumes a well-mixed setup, which actually corresponds to the individual scenario, where the CD-DC pairs contribute to the increase of cooperation while the CC--DD pairs do the opposite because the cooperators are at a disadvantage position, i.e. $\Pi^{G_{1,2}}_C-\Pi^{G_{1,2}}_D<0$.

 \begin{figure}[t]
\centering
\includegraphics[width=0.98\linewidth]{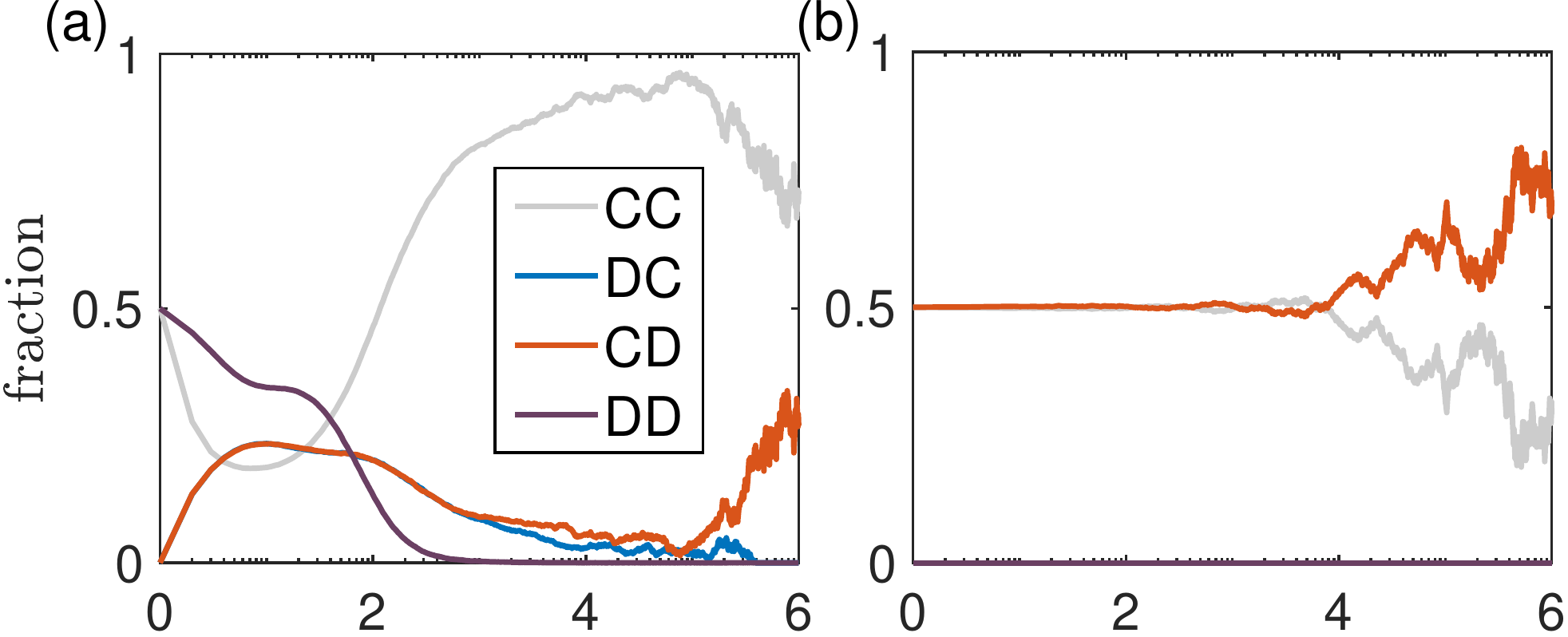}
\includegraphics[width=0.98\linewidth]{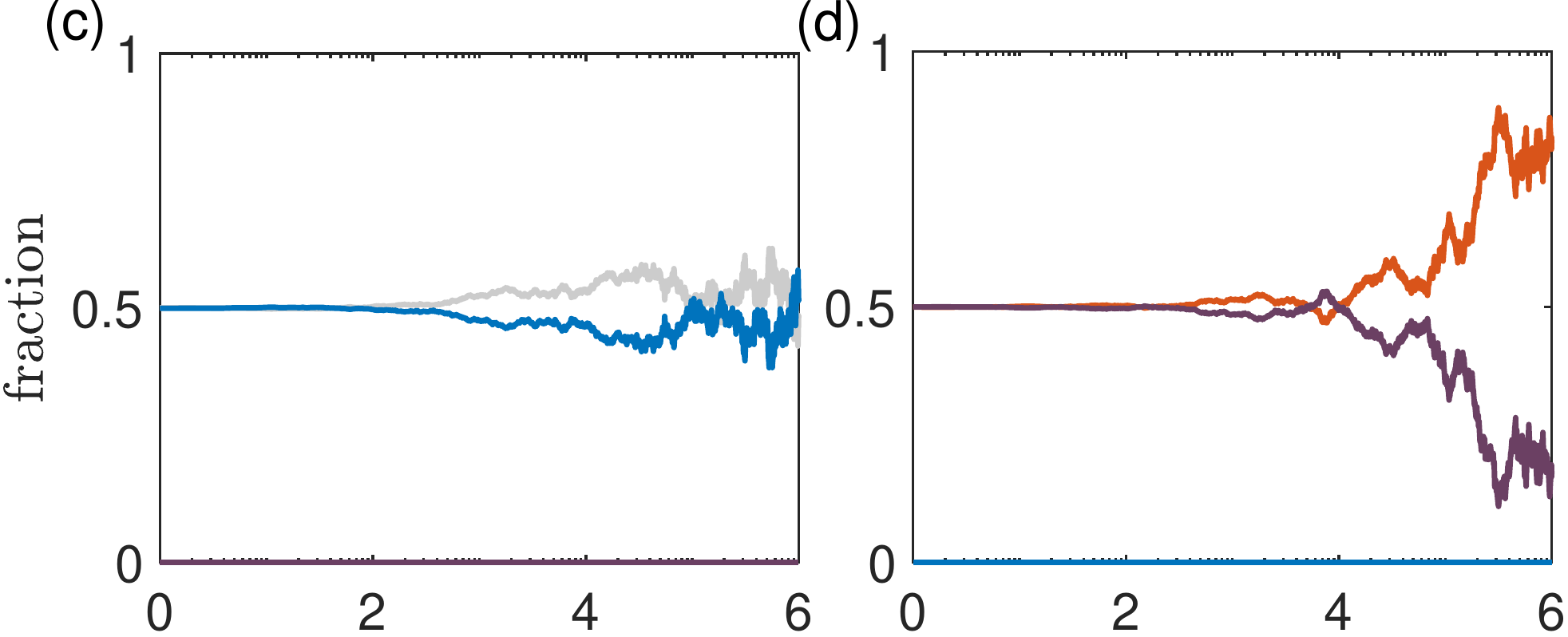}
\includegraphics[width=0.98\linewidth]{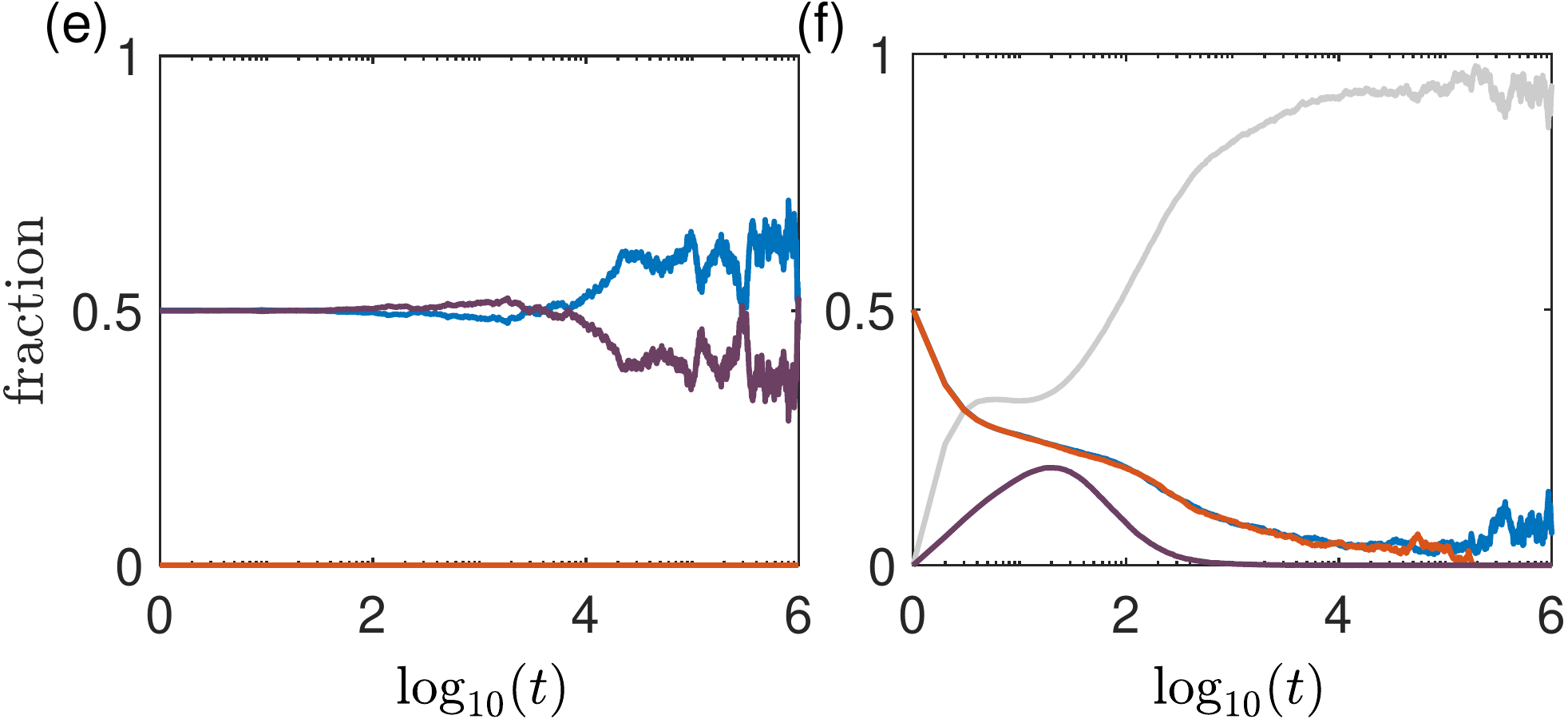}
\caption{(Color online)
Time series of all four fractions for two cross-played ($\theta=1$) PD games starting from random initial conditions.
All six binary compositions are included:
(a) CC--DD, (b) CC--CD, (c) CC--DC, (d) CD--DD, (e) DC--DD, (f) DC--DC. (a), (b-e), (f) correspond to category (i -- iii), respectively. 
Parameters: $S=0$, $T=1.1$, $R=1$, $P=0$, $L=1024$ for the 2d square lattice.
}
\label{fig:individual}
\end{figure}
 \begin{figure}[t]
\centering
\includegraphics[width=0.98\linewidth]{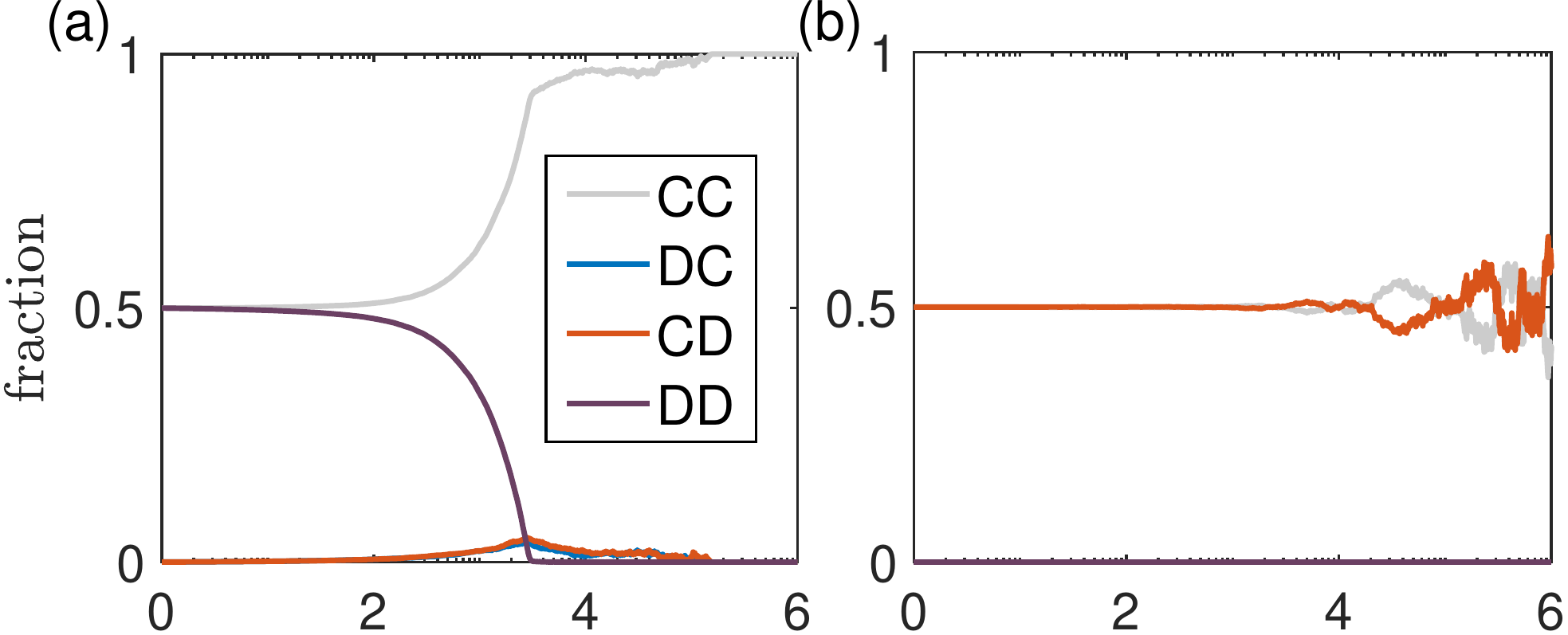}
\includegraphics[width=0.98\linewidth]{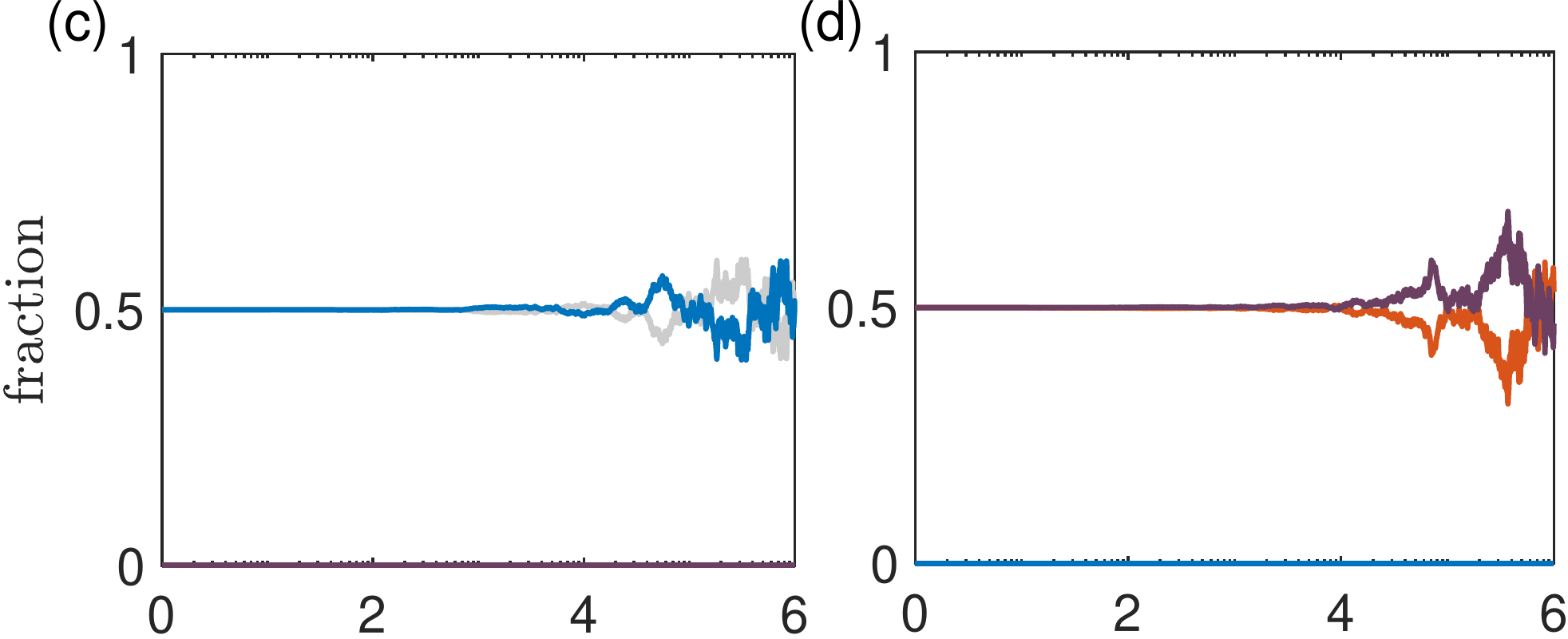}
\includegraphics[width=0.98\linewidth]{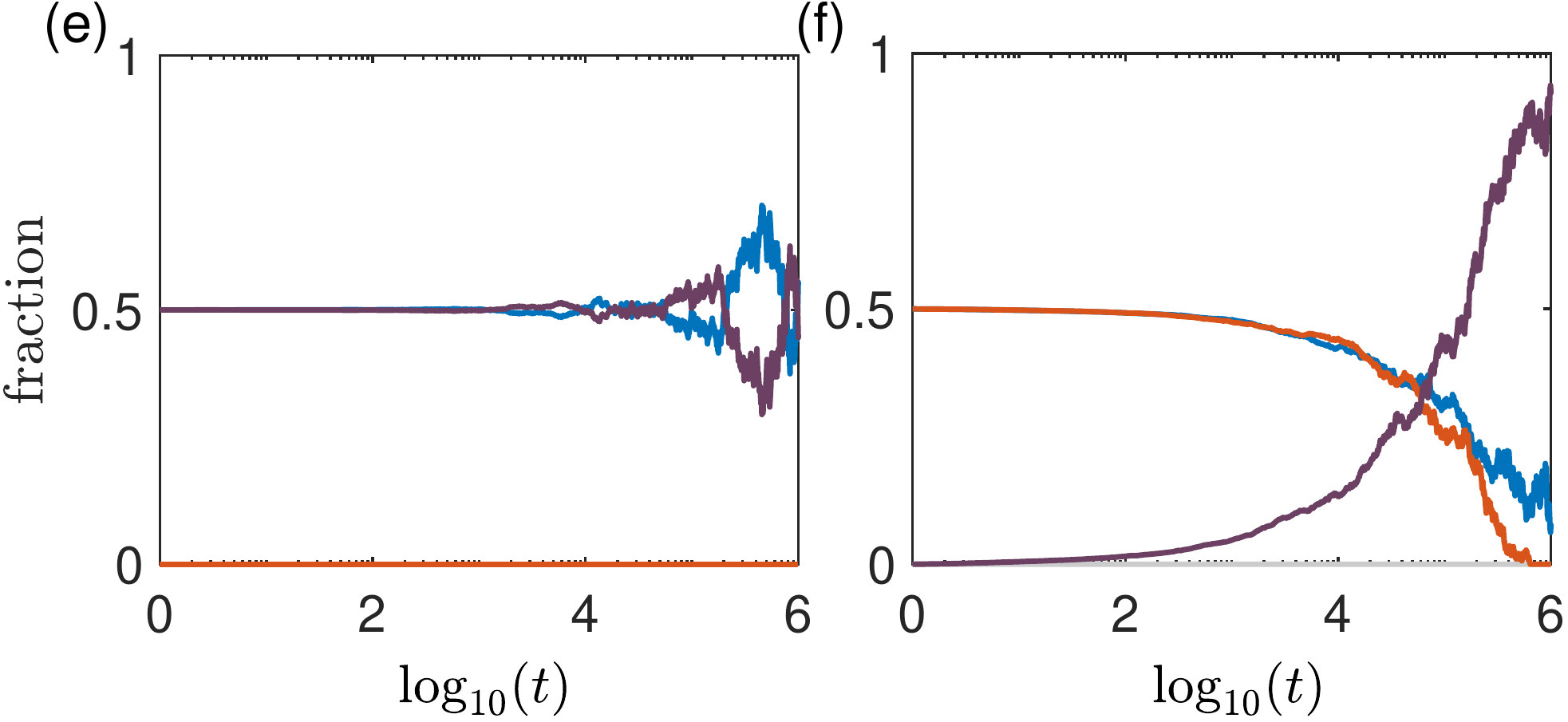}
\caption{(Color online)
Time series of all four fractions for two cross-played ($\theta=1$) PD games starting from half-half patched initial conditions. 
(a) CC--DD, (b) CC--CD, (c) CC--DC, (d) CD--DD, (e) DC--DD, (f) CD--DC. (a), (b-e), (f) correspond to category (i -- iii), respectively. 
Parameters: $S=0$, $T=1.1$, $R=1$, $P=0$, $L=1024$ for the 2d square lattice.
}
\label{fig:bulk}
\end{figure}
\subsection{Numerical evidences for the two scenarios}\label{subsec:evidence}
To validate the arguments of the three categories, some numerical experiments are conducted.  
To prepare the two scenarios, we respectively start with random and half-half patched initial conditions to mimic the individual and bulk circumstances respectively, and focus on the early stage of evolution, since longer evolution is going to ruin the randomness or compactness. Each time we only place two species on the 2d square lattice for clarity.

Figure~{\ref{fig:individual}} shows the time series for all six  binary combinations in the individual scenario. Category (i) corresponds to Fig.~{\ref{fig:individual}}(a), a mixture of CC and DD individuals. As can be seen, the fraction of CC decreases rapidly at the early few steps, DDs' density doesn't increase either,  instead the partial cooperators CD and DC show considerable increases. This is line with our argument that CC has no advantage against DD individually, and CD and DC are produced. The evolution at the later stage, the well-mixture is ruined and the CC players gain their strength to increase.
Figure~\ref{fig:individual}(b)-\ref{fig:individual}(e) reports the evolution of Category (ii). We observe that only the two prepared species are present, the others are of zero density. This is because a partial absorbing state is reached, e.g. CC--CD in Fig.~\ref{fig:individual}(b), game $G_1$ is in the absorbing state with full cooperation, the fraction of DC or DD continues to be zero. The remaining two fractions fluctuate around 1/2, the neutrality of their evolution is then confirmed.
Category (iii) is illustrated in Fig.~\ref{fig:individual}(f), where CD and DC players are randomly blended. As expected, in the first few steps, the fraction of CC rapidly increases together with some DD individuals, and the fractions of CD and DC themselves decease. Later on, the fraction of CC individuals continue to increase as all other three fractions decrease, but the population by then is not well-mixed anymore and some clusters are formed.

Figure~{\ref{fig:bulk}} shows the evolution in the bulk circumstance.
Category (i) is shown in Fig.~\ref{fig:bulk}(a), where the evolution of the two fractions is now qualitatively different from Fig.~\ref{fig:individual}(a) --- bulk CC individuals are now in an apparent advantage position over DDs, its fraction continues to increase and the faction of DD decreases until distinction. The fractions of CD and DC keep at relatively low densities. Typical spatiotemporal snapshots for this case are shown in Fig. 3 in CSL~\cite{CSL}. 
Category (ii) is illustrated in Fig.~\ref{fig:bulk}(b-e). As can be seen, the dynamical evolution is qualitatively the same as shown in Fig.~\ref{fig:individual}(b-e), all fluctuating around 1/2.  
Category (iii) is shown in Fig.~\ref{fig:bulk}(f), which is fundamentally different from the process in Fig.~\ref{fig:individual}(f), but in line with our arguments above. It shows that once CD and DC meet up in bulks, the payoffs from the intra-cluster gaming reverse the advantages, favoring the DD individuals.
As a result, the fractions of CD and DC continue to decrease and the density of DD monotonically increases without any CC individuals being seen.

Put together, the numerical experiments in Fig.~\ref{fig:individual} and Fig.~\ref{fig:bulk} perfectly justify the classification of the three categories.

\subsection{Dynamical realities on 2d square lattices}\label{subsec:reality}
However, these above two initial conditions are peculiar that correspond to two extreme circumstances and only last for a short-term time window. 
What really happened to these diverse interactions along the whole time evolution? Figuring out these facts is the key to understanding the working of dynamical reciprocity, since the invasion and catalyzed types of interaction always produce the opposite results, the classification itself cannot explain why cooperation is preferred.

\begin{figure}[t]
\centering
\includegraphics[width=0.99\linewidth]{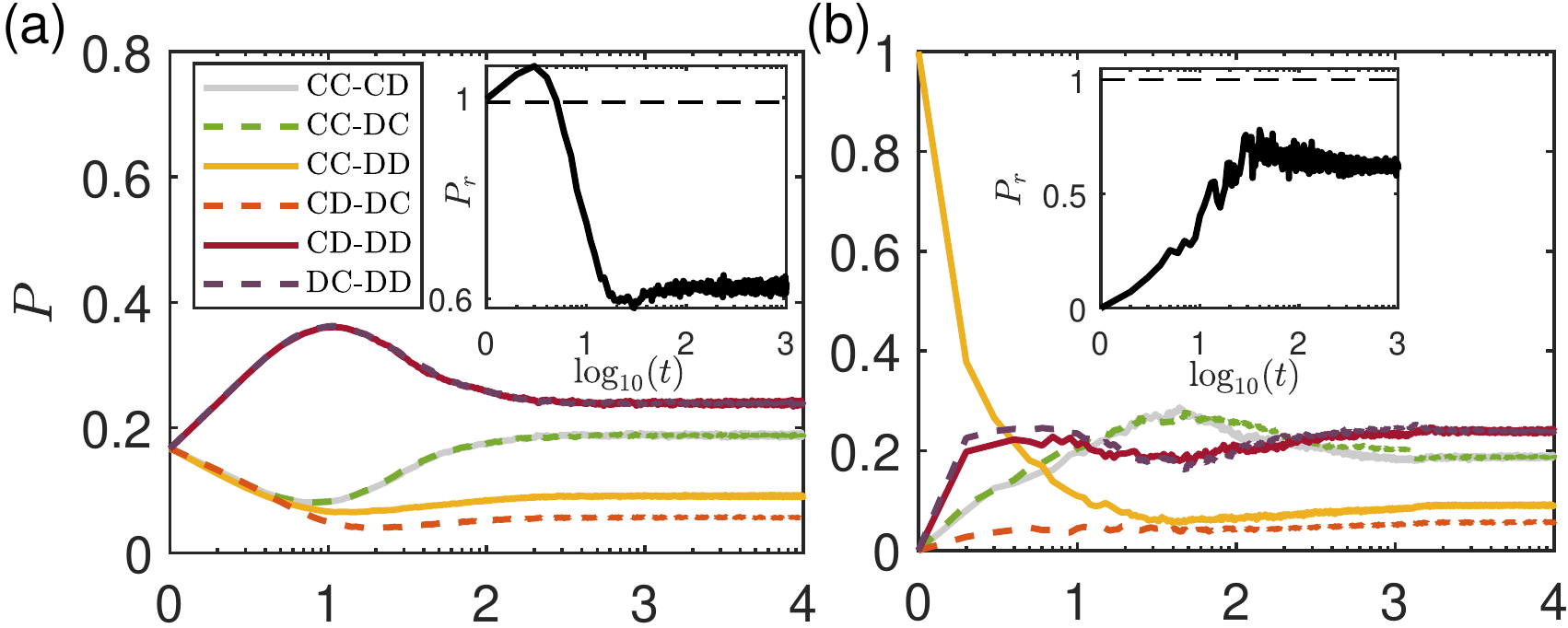}
\includegraphics[width=0.99\linewidth]{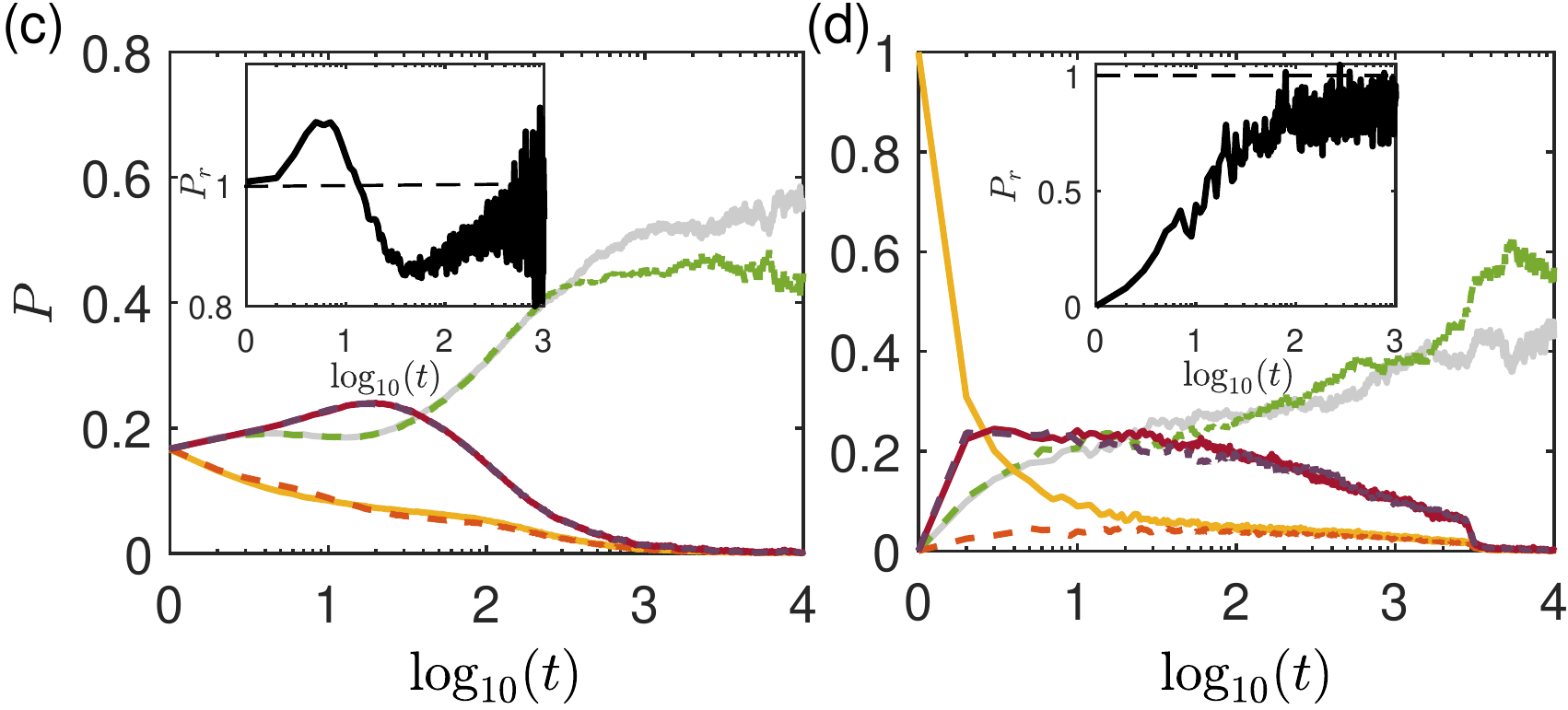}
\caption{(Color online)
Time evolution of interacting pair proportions for two interacting PD games.
Both random (a, c) and half-half patched (b, d) initial conditions are considered. 
(a, b) and (c, d) correspond $\theta=0.5$ and 1, respectively.
$P_r=P_{CD-DC}/P_{CC-DD}$ is to compare the relative proportion for CD--DC and CD--DD pairs.
Parameters: $S=0$, $T=1.1$, $R=1$, $P=0$, $L=1024$ for the 2d square lattice.
}
\label{fig:interface}
\end{figure}

Here, by adopting random initial conditions, we monitor the long-term evolution for both $\theta=0.5$ and 1, the dynamical processes typically experience two stages: 

i) At early stage $t<t_c$ ($t_c \sim 10$ MC steps for the size of $1024\times1024$), no sizeable clusters are supposed to be present in the system. Hence the individual interaction scenario applies. Statistically, as Fig.~\ref{fig:interface}(a,c) show that, there is a detectable quantity difference in CC--DD and CD--DC pairs that the proportion $P_{CD-DC}>P_{CC-DD}$, meaning the catalyzed interactions occur more frequently. Net production of cooperators is thus expected. Though this stage is relatively short.

ii) When $t>t_c$, clusters are gradually formed. Once the cluster property is strengthened, the bulk interaction scenario comes into play. Interestingly, a proportion crossover is found now that $P_{CD-DC}<P_{CC-DD}$, whereby cooperation is again enhanced since net cooperators are also favored in this scenario according to Table~\ref{tab:2games}.

Figure~\ref{fig:interface}(a,c) further show that the four neutral interacting pairs have the major proportions in most of the time for both cases, though they bring no net production of cooperators or defectors. The difference between Fig.~\ref{fig:interface}(a) and~\ref{fig:interface}(c) is that an equilibrium state of coexistence is reached for the case of $\theta=0.5$, while an absorbing state regarding $G_1$ is approached for $\theta=1$ where four interfaces except for CC--CD and CC--DC are vanishing.

Put together, for the whole processes, the system self-organizes into states with different relative proportions of invasion and catalyzed interaction type that make cooperators continuously be produced, no matter clusters are formed or not. 
Look back to the Eq. (\ref{eq:mf_cor}), we are now sure that it also captures the cooperation mechanism within the structured population. When the strategies are not clustered, defectors dominate i.e. $\Pi^{G_{1,2}}_C-\Pi^{G_{1,2}}_D<0$, but due to the number difference of the interacting pairs, $f_{CC}f_{DD}-f_{CD}f_{DC}<0$, the second term in Eq. (\ref{eq:mf_cor}) is thus positive. When the strategies are clustered, the opposite is true $\Pi^{G_{1,2}}_C-\Pi^{G_{1,2}}_D>0$, $f_{CC}f_{DD}-f_{CD}f_{DC}>0$, again the second term is positive and cooperation is enhanced. This means that the game-game interactions prefer cooperation in the whole evolutionary process and thus explains the dynamical reciprocity.

\begin{figure}[t]
\centering
\includegraphics[width=0.98\linewidth]{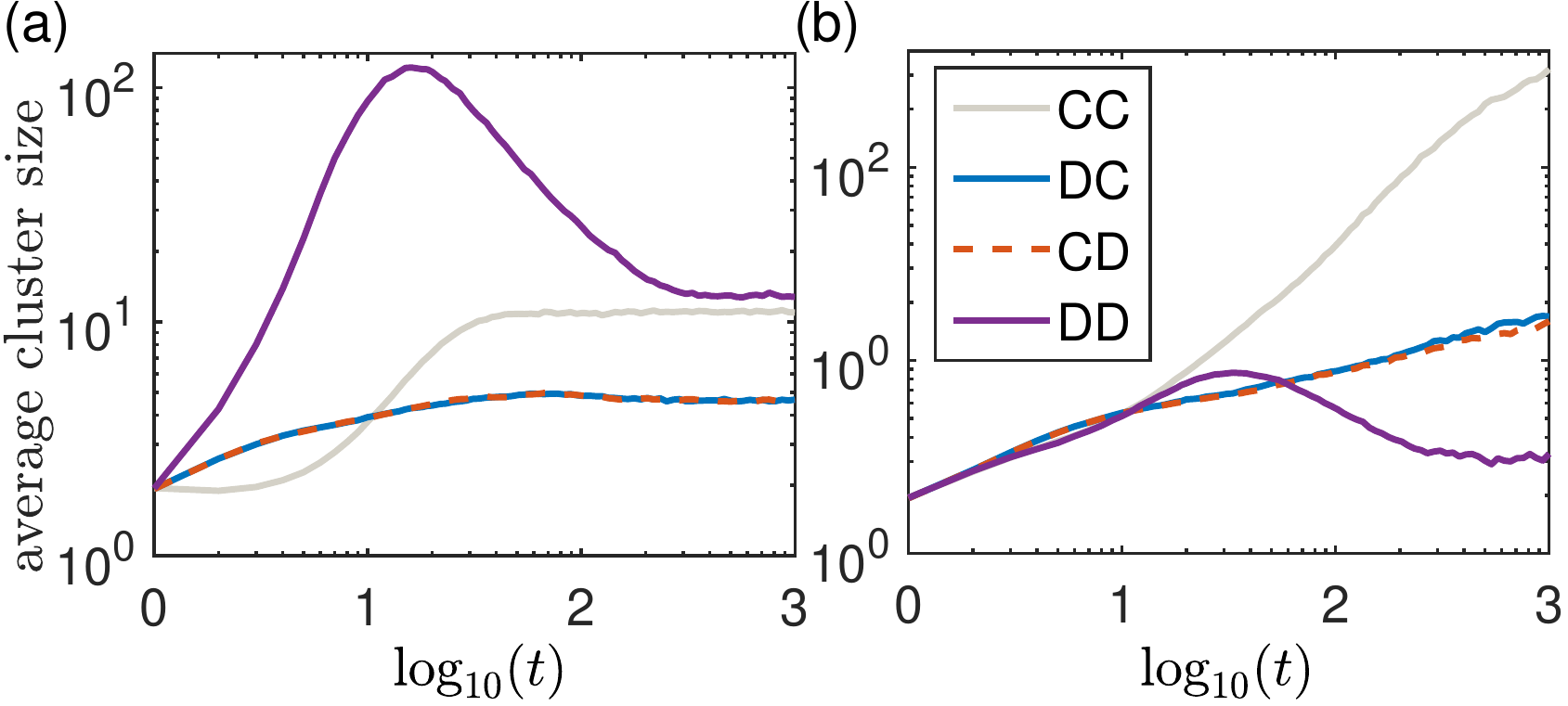}
\includegraphics[width=0.98\linewidth]{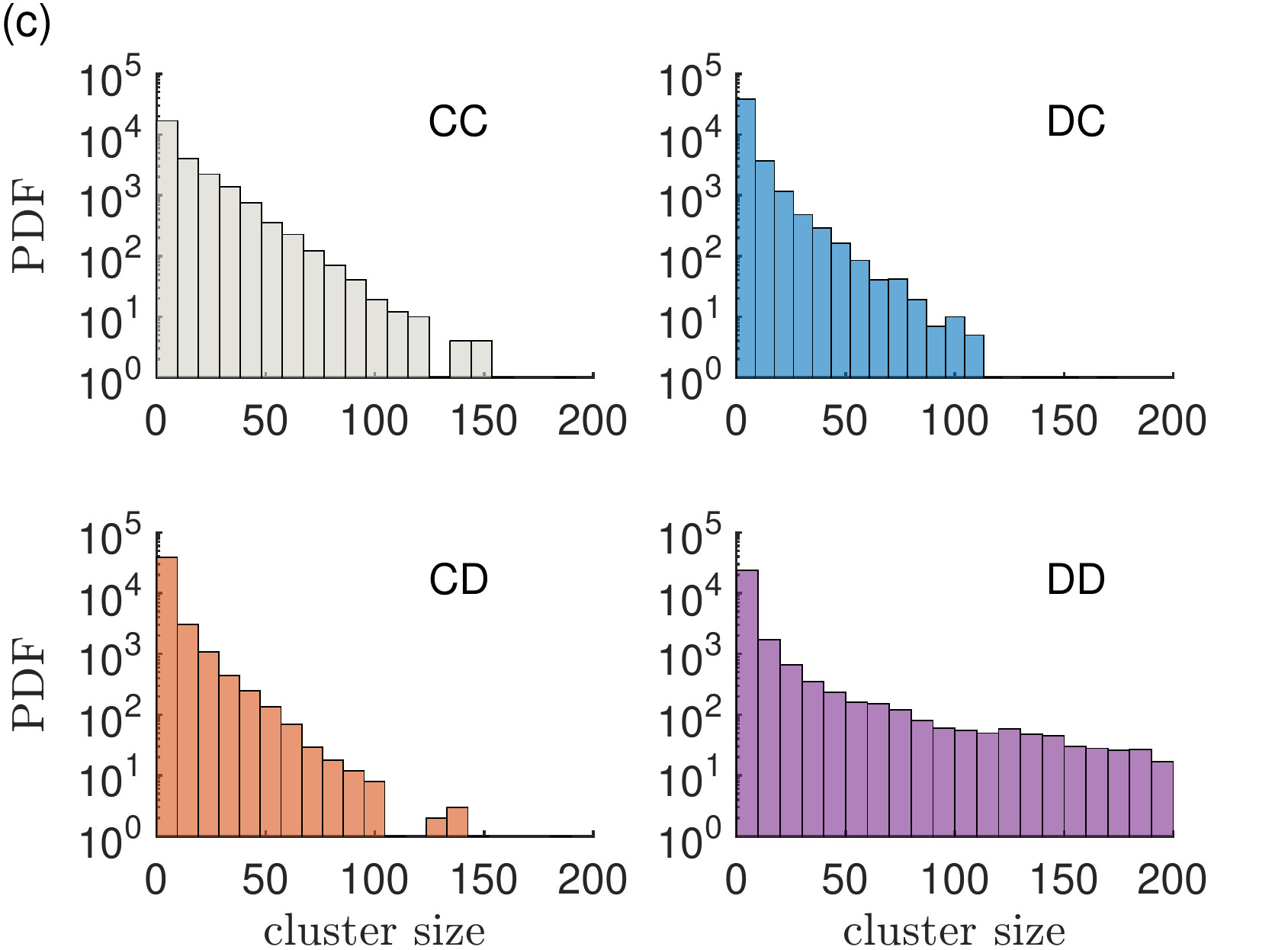}
\includegraphics[width=0.98\linewidth]{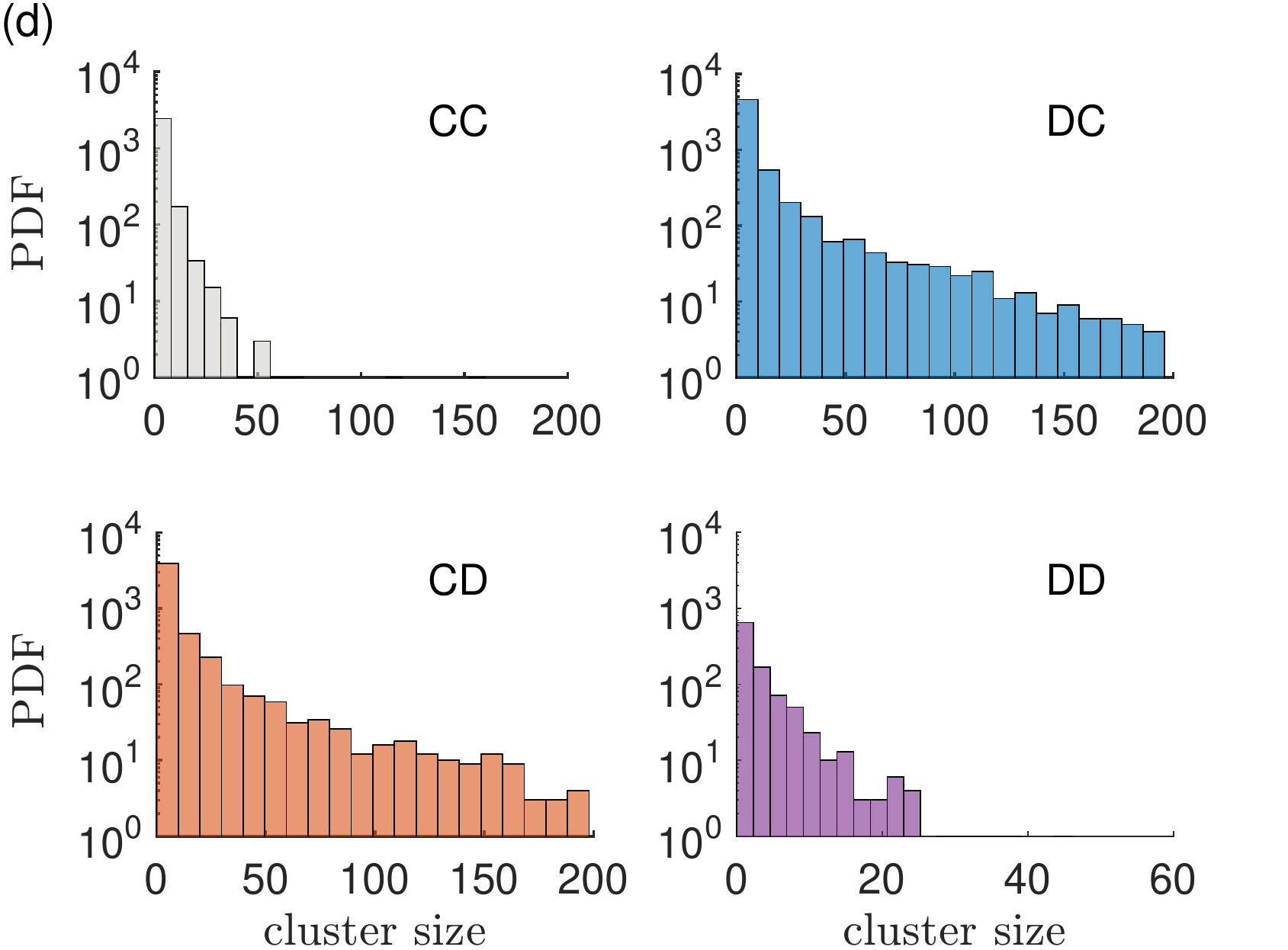}
\caption{(Color online)
Cluster size distribution for two interacting PD games starting from random initial conditions. 
(a, b) are time evolution of the average cluster size of the four species for $\theta=0.5$ and 1, respectively.
(c, d) are probability function distributions of cluster size at $t=1000$ for $\theta=0.5$ and 1, respectively. 
In (d), there is a giant CC cluster with the size comparable to the population size ($\sim N$), which is not shown.
Parameters: $S=0$, $T=1.1$, $R=1$, $P=0$, $L=1024$ for the 2d square lattice.
}
\label{fig:clustersize}
\end{figure}
 \begin{figure}[th]
\centering
\includegraphics[width=0.98\linewidth]{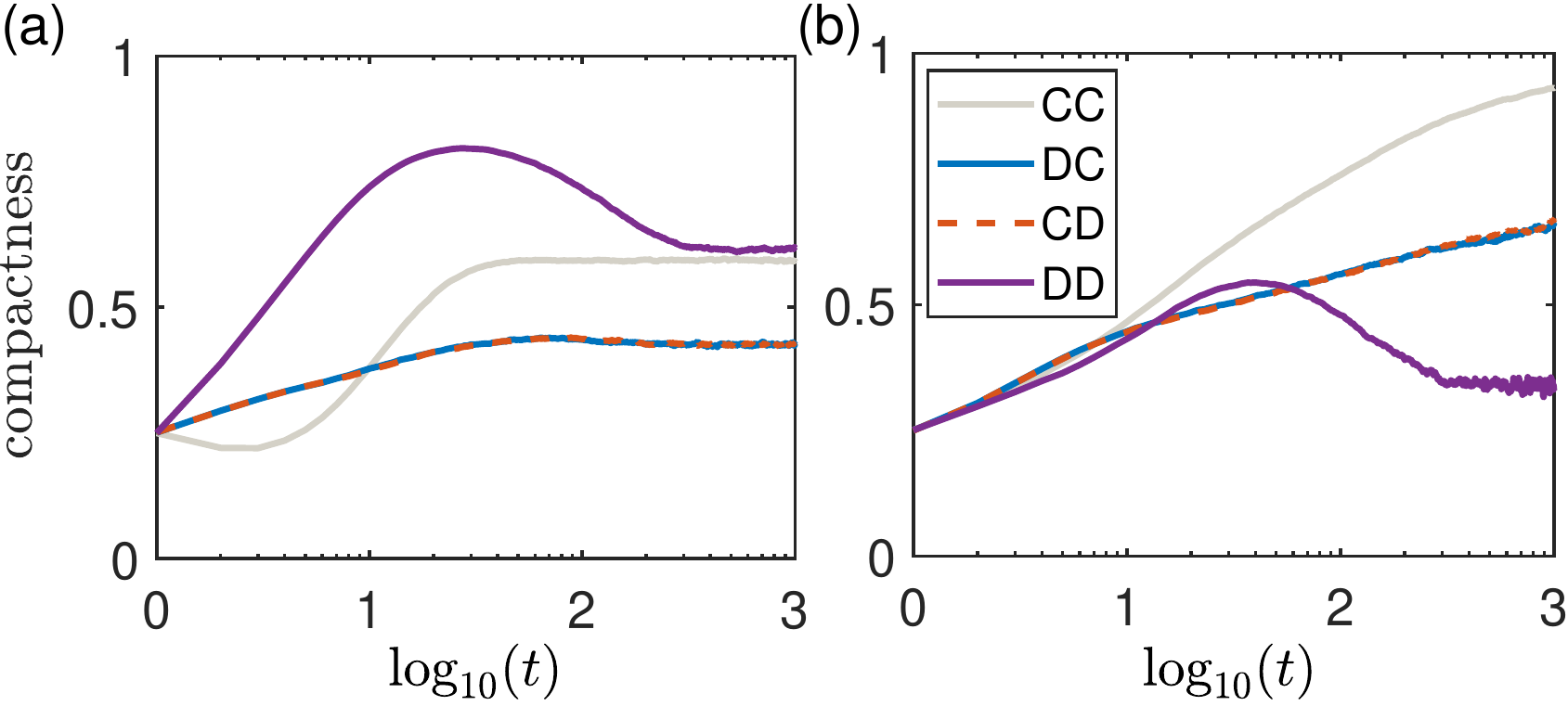}
\includegraphics[width=0.98\linewidth]{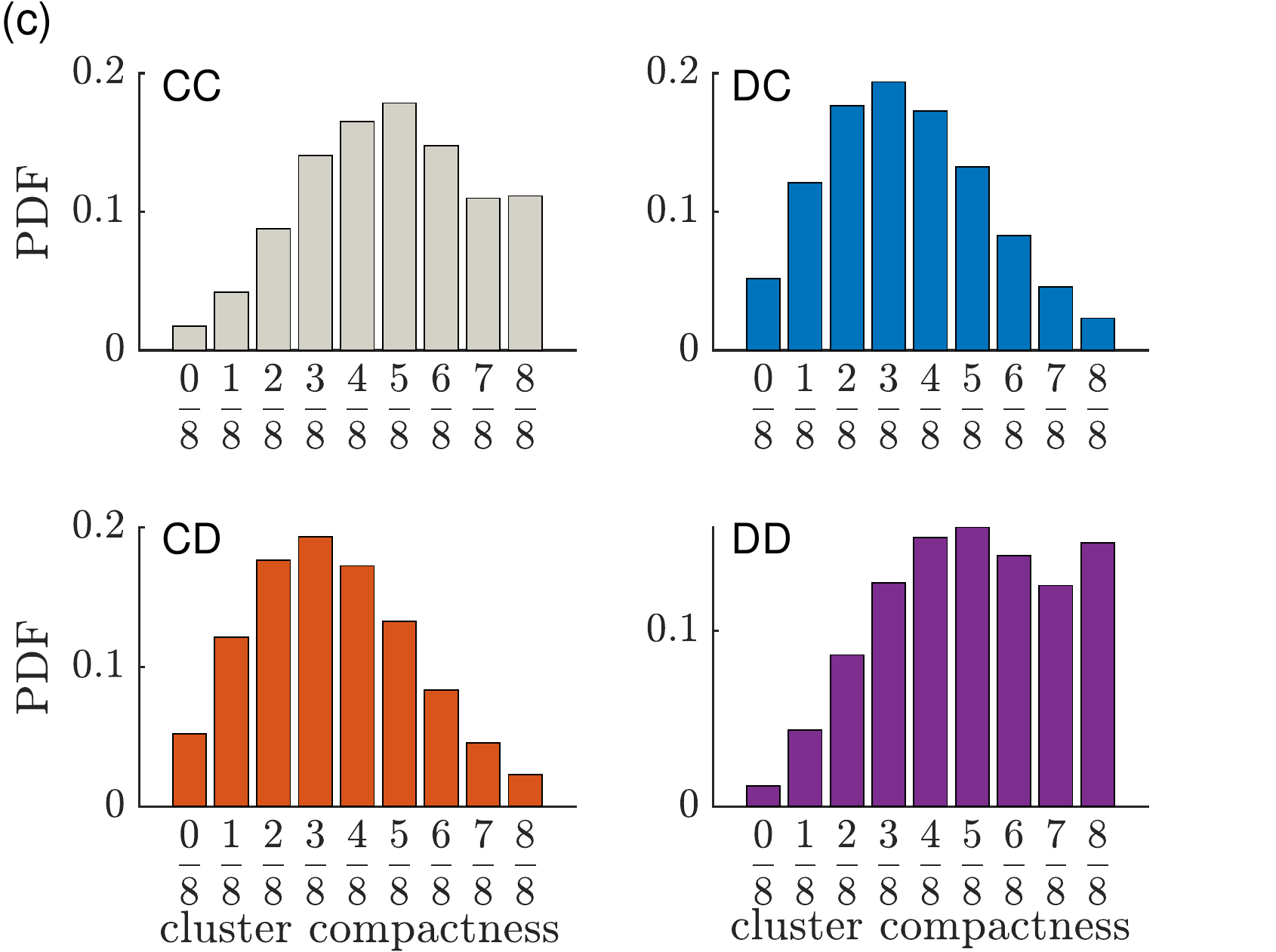}
\includegraphics[width=0.98\linewidth]{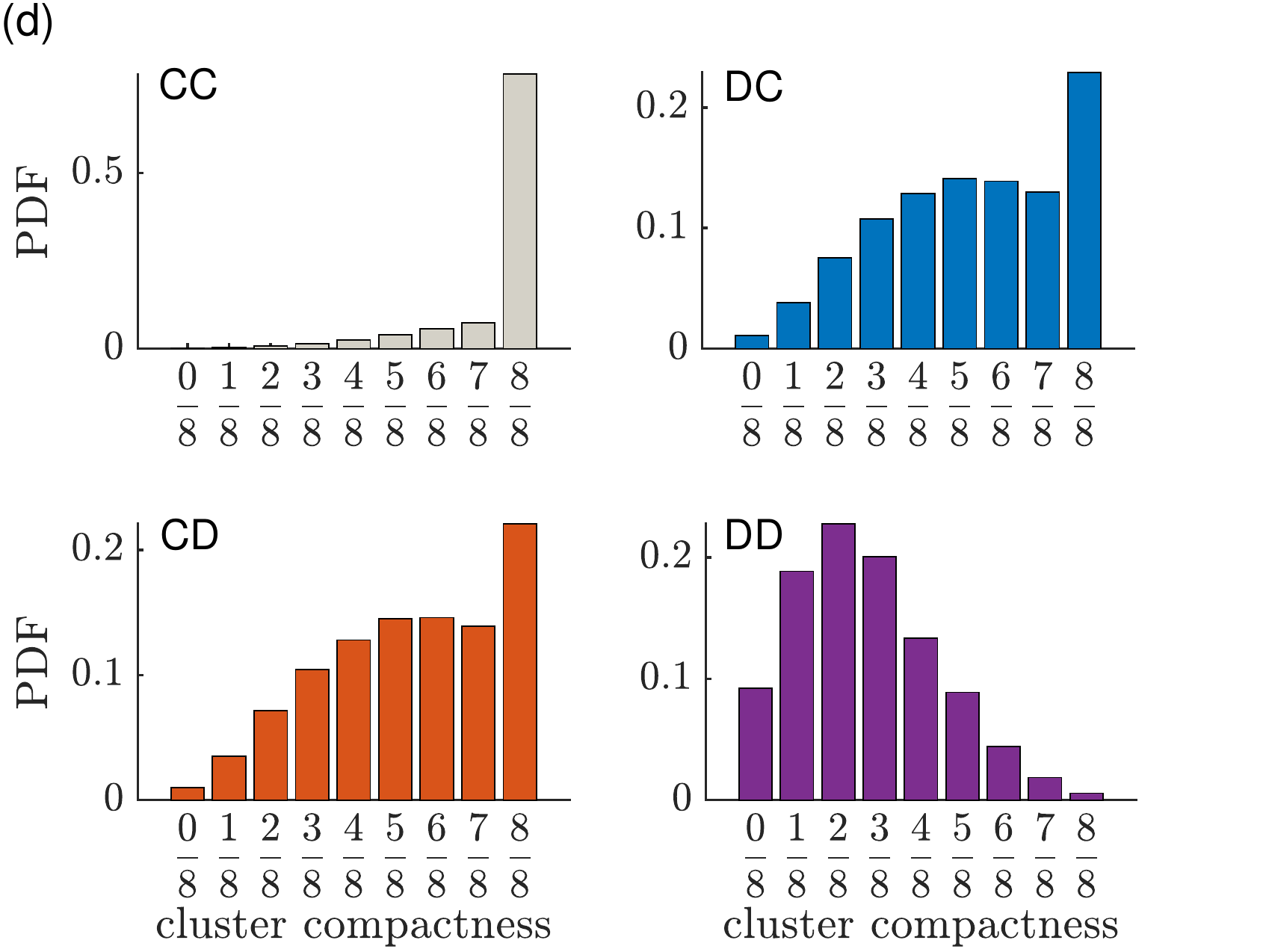}
\caption{(Color online)
Cluster compactness analysis. 
(a, b) Time evolution of compactness of the four species for $\theta=0.5$ and 1, respectively.
(c, d) Probability function distributions of cluster compactness at $t=1000$ for $\theta=0.5$ and 1, respectively. 
Same settings as in Fig.~\ref{fig:clustersize}.
}
\label{fig:compactness}
\end{figure}

A simpler case is starting from half-half patched initial conditions, where bulk scenario sets in from the very beginning, see Fig.~\ref{fig:interface}. We can see that the interface proportion of CC--DD decreases and all others increase, but the inequality $P_{CD-DC}<P_{CC-DD}$ holds along the whole process, no crossover as in Fig.~\ref{fig:interface}(a,c) is seen. The ensuing dynamics exhibits insensitivity to the initial conditions, the evolution is qualitatively the same as in Fig.~\ref{fig:interface}(a,c) in the long term. This means that in the absence of stage i), the dynamics in bulk scenario is still sufficient to yield high level of cooperation. 

\subsection{Cluster size and compactness analysis}\label{subsec:cluster}
To validate the appropriateness of the two-scenario division in the random initial condition case, we further conduct cluster size and compactness analysis. 
While the former is easy to understand and often adopted, the latter is to measure how compact of clusters, which is defined as the fraction of neighbors with identical state regarding the central player (here we adopt Moore neighborhood with eight neighbors), its average characterizes the overall compactness of clusters. The situation with both large average cluster size and compactness provides an ideal circumstance for bulk interaction scenario to play, the individual interaction scenario works for just the opposite case. There may also be cases that the average size is large but with small compactness or the way around, which are in between the two interaction scenarios discussed above.

Figure~\ref{fig:clustersize} provides the statistical properties of the cluster size for both $\theta=0.5$ and 1. As can be seen, the initial size of clusters is pretty small, but they become larger as time evolves, and get saturated in the case of $\theta=0.5$ at around $t\gtrsim t_c$ [Fig.~\ref{fig:clustersize}(a)]. For the cross-play case ($\theta=1$), however, the cluster sizes continue to increase except the DD players [Fig.~\ref{fig:clustersize}(b)]. The PDF shown in Fig.~\ref{fig:clustersize}(c,d) shows that the cluster size could typically reach an order of $10^2$ at the late stage but not for CC or DD clusters in the case of $\theta=1$, where they are either too huge or too small.

Figure~\ref{fig:compactness} shows the corresponding compactness analysis. Fig.~\ref{fig:compactness}(a,b) show similar profiles of time series when compare with Fig.~\ref{fig:clustersize}(a,b). Note that the peaks of DD species in both Fig.~\ref{fig:clustersize}(a,b) and Fig.~\ref{fig:compactness}(a,b) are simply due to the invasion of defectors at the initial stage for $t<t_c$. The PDF shows quite a few individuals are of high compactness, especially for the case of $\theta=1$, where there are some considerably large clusters (expect DD clusters) within the population.
These observations justified our the two-scenario division, whereby the above mechanism analysis of the dynamical reciprocity seems reasonable.

\section{Robustness studies}\label{sec:robustness}
In this section, we turn to provide more evidences to examine the robustness of the dynamical reciprocity by studying different variants of the above model, such as asymmetrically interacting games, games of the time-scale separation, and various updating rules etc. 
\begin{figure}[htbp]
\centering
\includegraphics[width=0.98\linewidth]{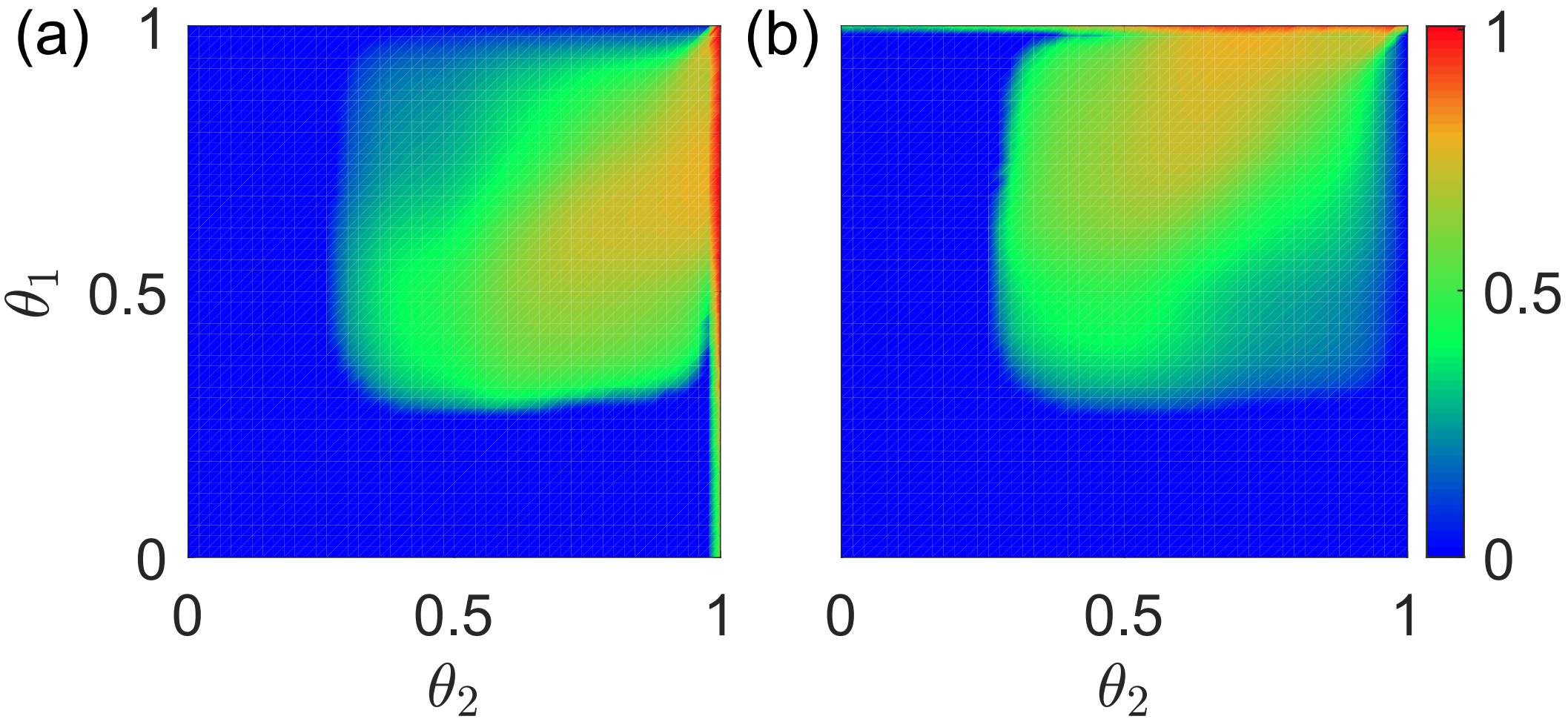}
\caption{(Color online)
Color coded cooperation prevalences for two asymmetrically interacting PD games on 2d square lattice within $\theta_1-\theta_2$ parameter space, (a) and (b) are for game $G_{1,2}$, respectively.
Other parameters: $S=0$, $T=1.1$, $R=1$, $P=0$, $L=128$.
}
\label{fig:asymmetry}
\end{figure}
\subsection{Asymmetrically interacting games}
The first relaxation of the above model is to remove the symmetry assumption by using asymmetric settings, which could be more realistic.
One variant could be that the two games are identical, while their impact on each other is assumed to be asymmetric. The effective payoffs defined in Eq.~(\ref{eq:effective_linear}) are then naturally written as
\begin{equation}
\begin{pmatrix}
\widehat\Pi^{G_1}_{x} \\ \widehat\Pi^{G_2}_{x} 
\end{pmatrix}
=
\begin{pmatrix}
1\!-\!\theta_2 & \theta_2 \\
\theta_1 & 1\!-\!\theta_1\!  
\end{pmatrix}
\begin{pmatrix}
\Pi^{G_1}_{x} \\ \Pi^{G_2}_{x} 
\end{pmatrix},
\label{eq:asy}
\end{equation}
where the interaction strengths $\theta_{1,2}\in[0,1]$ represent the contribution fraction of game $G_{1,2}$ in the other game's effective payoff.
If a game is more important for the other game than the way around, then generally $\theta_1\!\neq\! \theta_2$. Note that $\theta_{1,2}$ are not necessarily negatively or positively correlated, both could view the other game important or unimportant in their decision-makings, up to the specific context. 

Figure~{\ref{fig:asymmetry}}(a,b) show respectively the cooperation levels for the two games within the $\theta_1 - \theta_2$ parameter space, where the symmetrical case studied above is along $\theta_1\!=\!\theta_2$. The overall trend is qualitatively the same as the symmetrical case that a high level of cooperation is expected when $\theta_{1,2}$ become large. In particular, a high cooperation level of a given game, say game $G_1$,  is more likely to happen when the other game's contribution $\theta_2$ is large given its own's contribution is not too small ($\theta_1\gtrsim 0.4$); and nearly full cooperation is reached when $\theta_2$ approaches one. But due to the asymmetry, however, the two cooperation prevalences could be very different.

\begin{figure}[htbp]
\centering
\includegraphics[width=0.98\linewidth]{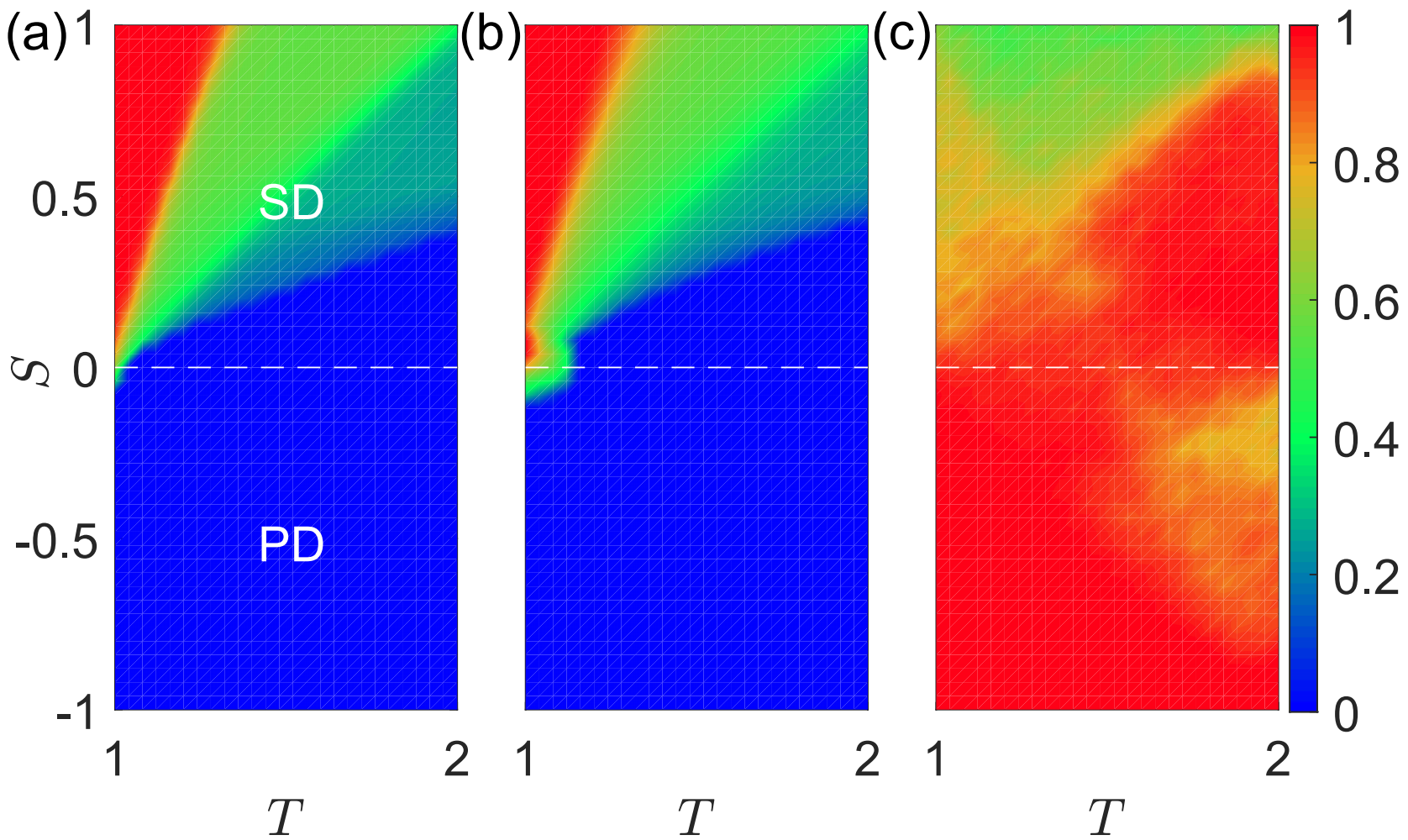}
\caption{(Color online)
Color-coded cooperation prevalence in $S-T$ parameter space for two interacting games --- one is PD and the other is SD, on the 2d square lattice, for $\theta=0, 0.5, 1$ (a-c). They have opposite sign in $S$, e.g. the SD at top left corner $(T=1,S=1)$ interacts with PD with $(T=1,S=-1)$ at the bottom left. 
Other parameters: $R=1$, $P=0$, and $L=128$ for the lattice.
}
\label{fig:PD_SD}
\end{figure}

\subsection{Interacting games composed of two different types}\label{subsec:PD_SD}
Another variant is to remove the symmetry in the two games per se.
Interacting games in the real world more often involve different issues, which should be modeled by different types of game.  Thus, here we study the case of interacting games composed of two different pairwise games, i.e. one is PD game, the other is SD, but with the same interaction strength $\theta$.  

To reduce the parameters, a convenient way to do is as follows~\cite{santos2014biased,wang2014evolutionary}: the two games share three identical parameters $R$, $P$, $T$, but with opposite signs in $S$, since $0<S<1$ for SD and $-1<S<0$ for PD are required by definition. Figure~\ref{fig:PD_SD} shows three typical interaction strengths in $S-T$ parameter space. Without game interaction, they are independent, almost no cooperation is expected for PD. As the interaction becomes stronger, both cooperation levels are promoted. Significant promotion is possible when the scenario becomes cross-playing, again high cooperation prevalences arise for the whole parameter space. Due to the difference of the two games, now the cooperation levels are not homogeneous, but with certain nontrivial distribution in the parameter space, remaining for further investigation.    

Note that here the game-game interaction promotes cooperation in both games, which is superior to the previous results within the framework of interdependent network. In ~\cite{santos2014biased,wang2014evolutionary}, the PD and SD are placed on two interdependent networks, and they are coupled through network interdependency.  They show that with the increment of coupling, the cooperation promotion in PD is at the expense of cooperation decline in SD.

\subsection{Interacting games with time-scale separation}\label{subsec:tss}
An underlying assumption in the above studies is that the time-scales of involved games are comparable, their updating rates are assumed to be identical. A further relaxed condition is to allow for different time scales, or even time-scale separation --- some of them are fast games while others are slow ones. This could be the case in reality since different issues by nature has its own paces and hence is reasonably of different time scales.

To see the impact of the time-scale separation, we study two interacting PD games, one is fast, the other is slow. The time-scale separation is characterized by the time scale ratio $T_r\geq1$: when the slow game is updated by one generation for each player, the fast game is updated $T_r$ generations on average. Apparently, the case of $T_r=1$ is reduced to identical time scale scenario as studied above; and as $T_r$ increases, the time-scale separation becomes stronger.

\begin{figure}[t]
\centering
\includegraphics[width=0.98\linewidth]{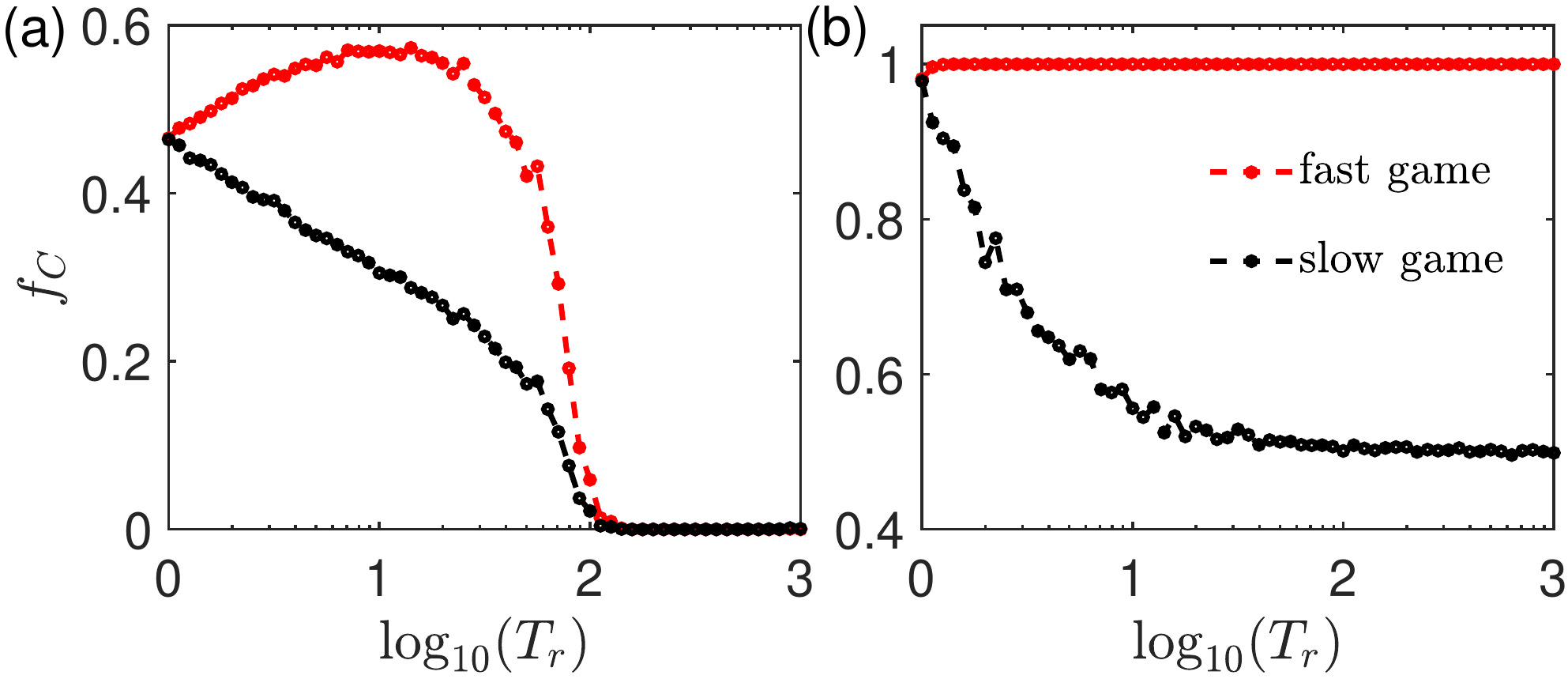}
\caption{(Color online)
The impact of time-scale separation of the two interacting PD games.
The cooperation prevalence $f_c$ versus the time-scale ratio $T_r$ for $\theta=0.5$ (a) and $\theta=1$ (b). 
No cooperation is seen for $\theta=0$ for the given parameters.
Other parameters: $S=0$, $T=1.1$, $R=1$, $P=0$, $L=1024$ for the 2d square lattice.
}
\label{fig:TSS}
\end{figure}

Figure~\ref{fig:TSS}  shows the cooperation prevalence for a wide range of time-scale separation for $\theta=0.5$ and 1.
In both cases, the cooperation prevalence in the slow game monotonically declines, while the cooperation level for fast game keeps largely unchanged or even increases within $T_r<10$ in the case of $\theta=0.5$. As $T_r$ further increases, the declines of both games are observed in Fig.~\ref{fig:TSS}(a), while the cooperation level for slow game keeps round 0.5 and full cooperation for the fast game in Fig.~\ref{fig:TSS}(b). These results mean that  a fairly high level of cooperation is still able to be maintained when the time scales are not very much separated. A systematic account of time-scale separation's impact will be presented elsewhere.

\begin{figure*}
\centering
\includegraphics[width=0.8\linewidth]{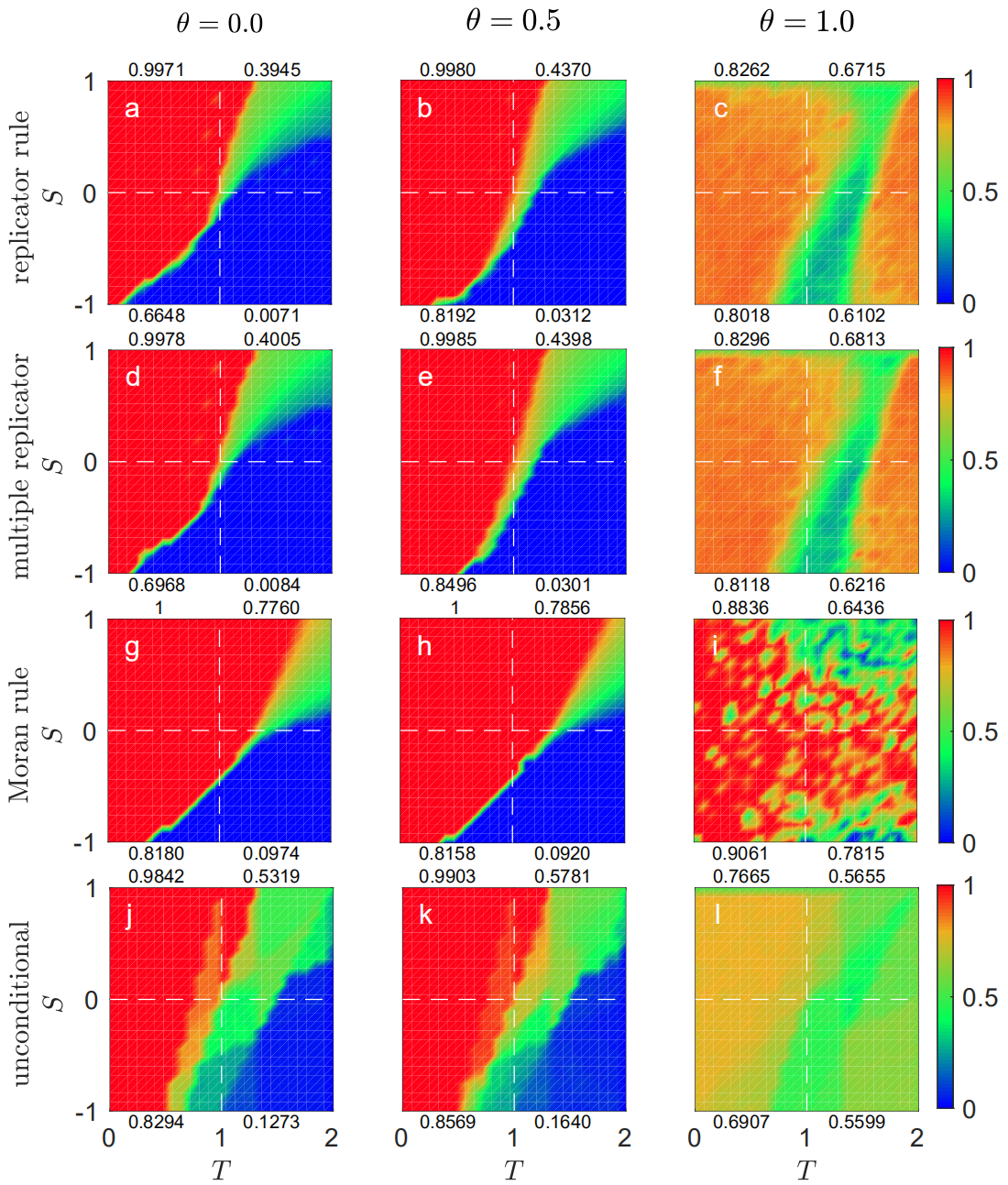}
\caption{(Color online)
Cooperation phase diagram for $f^{G_1}_C$ with other four update rules with AU, for $\theta=0,0.5,1$. 
The update rules are: replicator rule (first row, a-c), multiple replicator rule (second row, d-f), Moran rule (third row, g-i), and unconditional imitation (bottom row, j-l). 
Random initial conditions are adopted. The four numbers are the average cooperation prevalence for the corresponding quadrants.
Also $f^{G_2}_C\approx f^{G_1}_C$ due to the symmetrical settings.
Parameters: $R=1$, $P=0$, and $L=128$ for the 2d square lattice.
}
\label{fig:async}
\end{figure*}

\begin{figure*}
\centering
\includegraphics[width=0.8\linewidth]{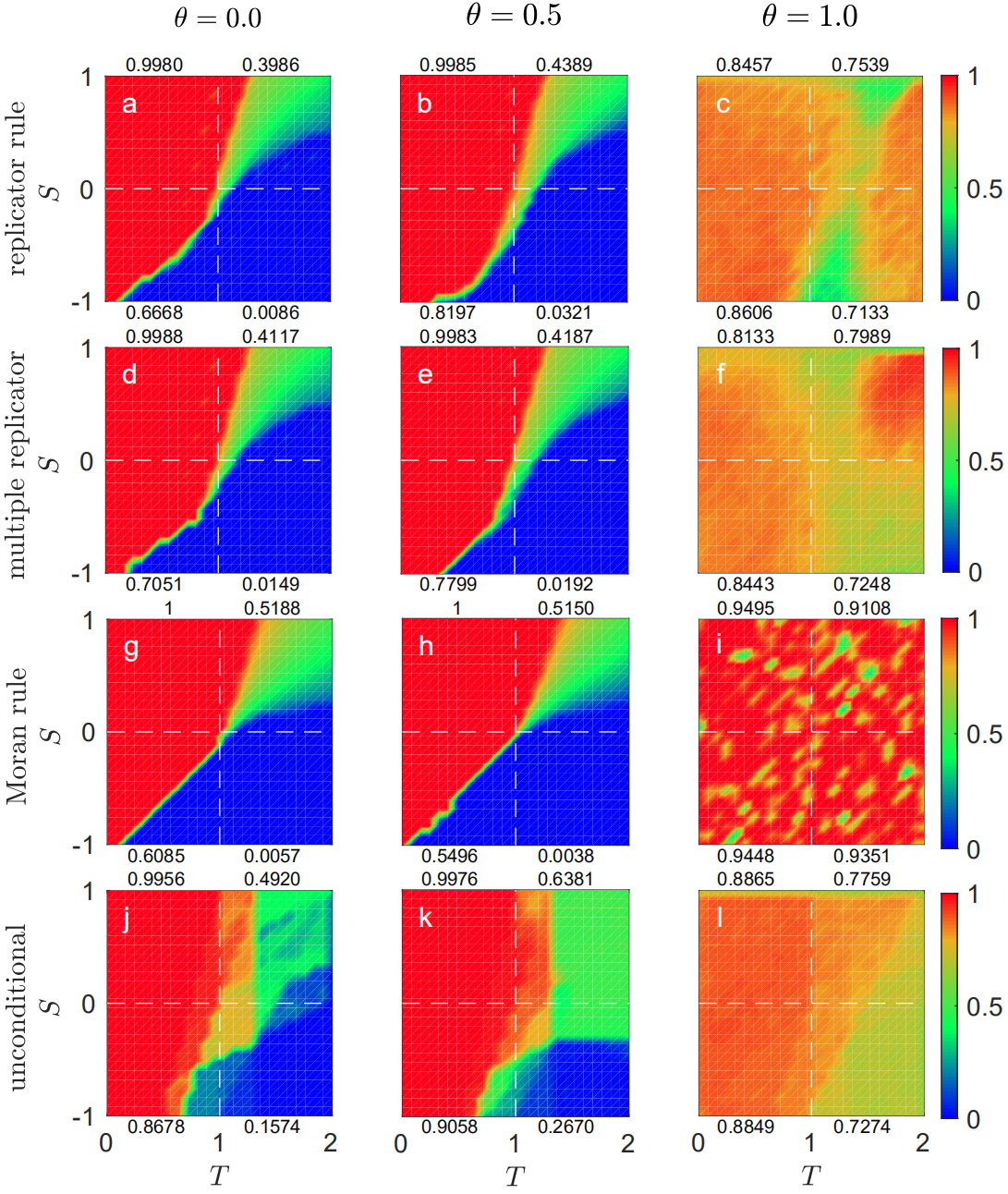}
\caption{(Color online)
Cooperation phase diagram with other four update rules but with SU, for three interaction strengths $\theta=0,0.5,1$. 
Other settings are exactly the same as in Fig.~\ref{fig:async}.
}
\label{fig:sync}
\end{figure*}

\subsection{Different updating rules and sychronicities}\label{subsec:rules}
The update rule determines how exactly the strategy of individuals evolves in time. There is a variety of update rules that have been adopted in the literature, each being conceived in different backgrounds, and sometimes they yield fairly different cooperation outcomes. For completeness, here we also examine four other often used updating rules \cite{roca2009evolutionary}:

(i) \emph{Replicator rule}, also known as the \emph{proportional imitation rule}, is inspired by the replicator dynamics. The procedure is similar as the Fermi rule case that we randomly pick one node $x$ and one of its neighbor $y$ and the imitation probability linearly depends on the payoff difference as:
 \begin{equation}
     W^g_{xy}\equiv W(s^g_y \rightarrow s^g_x)=\left\{
                \begin{array}{ll}
                  (\widehat{\Pi}_y^g - \widehat{\Pi}_x^g)/\widehat{\Pi}_0^g, \qquad \widehat{\Pi}_y^g>\widehat{\Pi}_x^g,\\
                  0, \qquad\qquad\quad\qquad  \widehat{\Pi}_y^g\leq\widehat{\Pi}_x^g,
                \end{array}
              \right. \label{eq:RR}
\end{equation}
where $\widehat{\Pi}_0^g=k(\max(1,T) - \min(0,S))$ to ensure the probability $W^g_{xy}\in[0,1]$.

(ii) \emph{Multiple replicator rule} is a variation of the replicator rule, where we now check simultaneously the whole neighborhood of $x$, and therefore it's more probable to change the strategy of $x$. With this rule, the probability of player $x$ maintaining its strategy is 
 \begin{equation}
W(s^g_x \rightarrow s^g_x)=\prod_{y\in\Omega_x} (1-W^g_{xy}),\label{eq:MRR}
\end{equation}
where $W^g_{ij}$ is given by (\ref{eq:RR}). The more neighbors an individual has, the less likely to maintain its strategy.

(iii) \emph{Moran-like rule}, also know as \emph{death-birth rule}, is inspired by the Moran process in biology. With this rule, an individual randomly picks one site in its neighborhood including itself, with the imitation probability defined as
 \begin{equation}
W(s^g_y \rightarrow s^g_x)=\frac{\widehat{\Pi}_y^g-\widehat{\Pi}_0^g}{\sum_{i\in \Omega^*_x} (\widehat{\Pi}_i^g-\widehat{\Pi}_0^g)},\label{eq:MLR}
\end{equation}
where $\Omega^*_x=\Omega_x \cup \{x\}$, the constant $\widehat{\Pi}_0^g$ is to guarantee the numerator positive, $\widehat{\Pi}_0^g=\max_{j\in\Omega^*_x} (k_j)\min(0,S)$ for pairwise games.

(iv) \emph{Unconditional imitation rule}, also known as \emph{follow-the-best rule}, is a deterministic rule. At each time step, every player adopts the strategy of the individual who has the highest payoff in its neighborhood, given this payoff is greater than its own.
 
Another complication of the model study is the synchronicity of the strategy update. As described in Sec.~\ref{sec:model}, the one used above is the random sequential updating, or simply termed as \emph{asynchronous updating} (AU). We also examine \emph{synchronous updating} (SU), where each individual is updated simultaneously, thus each player is updated exactly once per generation. Previous studies show that the synchronicity issue is often relevant for the cooperation outcome \cite{roca2009evolutionary}. In what follows, we investigate the cooperation dynamics of two interacting PD games using the above four updating rules, with both AU and SU. 

Figure~\ref{fig:async} presents the cooperation prevalence of two interacting PD games for the above four updating rules, for AU.  Without game interaction, the phase diagram are very similar, except the unconditional imitation rule case [Fig.~\ref{fig:async}(j)], where its phase diagram differs significantly from the others. As game-game interaction is engaged and becomes stronger, the overall cooperation is promoted in general, and this promotion reaches maximal as $\theta \rightarrow1$, the cross-playing scenario, in line with the Fermi rule studied above. In this scenario, however, there are some new dynamical features emerging. The most significant distinction is that  there is no monotonic dependence of cooperation on the temptation $T$, an intermediate value of $T$ yields a cooperation valley for replicator rule, multiple replicator rule and unconditional imitation [Fig.~\ref{fig:async}(c,f,l)]. Besides, many ``defection islands" arise within the parameter space for the the Moran rule [Fig.~\ref{fig:async}(i)]. Furthermore, the cooperation prevalence in some quadrants instead decreases, like HG and SH.

SU doesn't alter the overall phase diagram of cooperation, as shown in Fig.~\ref{fig:sync}. The cooperative behaviors are qualitatively the same as the cases with AU, and values are even slightly larger compared to Fig.~\ref{fig:async}, where the monotonic dependence of $T$ and ``defection islands" in the cross-playing scenario remain. Detailed analysis requires inspecting the evolutionary game dynamics, which will be presented elsewhere. 

Taken together, the observations in Fig.~\ref{fig:async}  and Fig.~\ref{fig:sync} mean that the dynamical reciprocity is robust against different rules and the synchronicity of the strategy updating. 

\section{Interacting games on complex networks}\label{sec:networks}
While the above robustness studies focus on the variations in dynamical aspects, we now turn to the impact of structural complexities, since the underlying connectivities of real populations are far more complex than regular lattices we studied. The stylized models for complex networks include Erd\H{o}s--R\'enyi (ER) random networks, small-world (SW) networks, scale-free (SF) networks, the impact of which on cooperation in the single game case has been studied extensively \cite{szabo2007evolutionary}, both theoretically and experimentally. Here we are not aiming to examine exhaustedly different topologies, rather we only study three of them and focus on the question: \emph{whether the structural complexities alter qualitatively the working of dynamical reciprocity?}

\begin{figure}[t]
\centering
\includegraphics[width=0.99\linewidth]{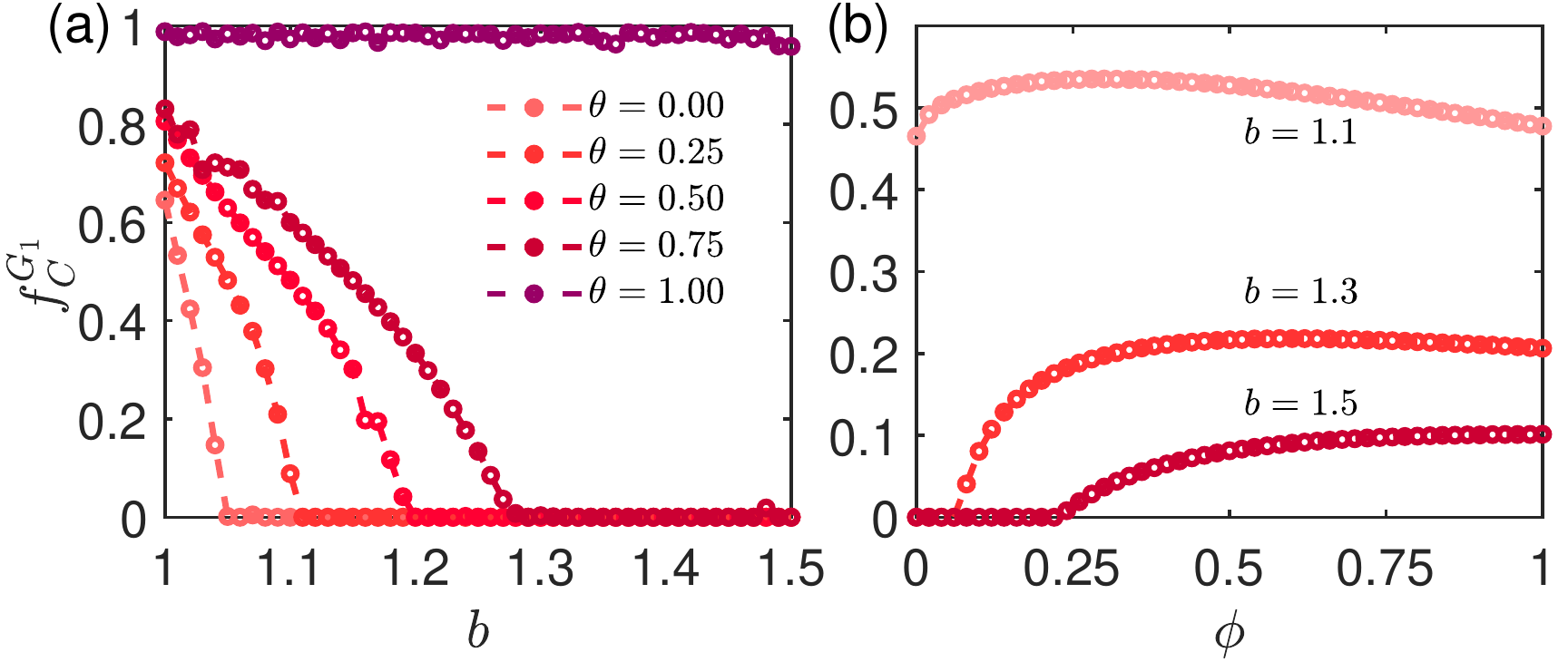}
\caption{(Color online)
Two interacting PD games on Newman-Watts small-world networks.
(a) The cooperation prevalence $f_c$ versus the temptation $T=b$ for a couple of interaction strengths $\theta$. 
(b) $f_c$ versus $\phi$ for three temptation values $b$ with $\theta=0.5$.
$f^{G_2}_C\approx f^{G_1}_C$ due to the symmetrical settings.
Parameters: $P=S=0$, $R=1$ for both games, the network size $N=2^{20}$ with $\phi=0.01$.
}
\label{fig:NW}
\end{figure}
\subsection{Small-world networks}\label{subsec:SW}
We adopt Newman--Watts network for the SW networks ~\cite{newman1999scaling}, which are derived from $d$-dimensional square lattice (here $d=2$) by adding some additional random links. This SW model is thought to be better behaved than the original network model~\cite{watts1998collective}, such as the exclusion of detached possibility. Instead of rewiring, shortcuts are added with a probability $\phi$ corresponding to each bond of original lattice, so that there are $dN\phi$ shortcuts on average. The average degree is then $\left\langle k\right\rangle=2d(1+\phi)$. By tuning the parameter $\phi$, the topology can vary continuously from the regular lattice to small-world networks, and to random networks in principle.

Fig.~\ref{fig:NW}(a) provides phase transitions for a couple of interaction strength $\theta$ for $\phi=0.01$, showing that as $\theta$ increases, the cooperation prevalence is continuously promoted; when $\theta\rightarrow1$, the phase transition disappears, and a fairly high level of cooperation is also observed, irrespective of the control parameter $b$. Fig.~\ref{fig:NW}(b) shows the cooperation dependence on the network parameter $\phi$ for three $b$. While a higher temptation $b$ reduces the cooperation prevalence as expected, a larger $\phi$ generally enhances the cooperation for large $T$. However, when the temptation is small (e.g. $b=1.1$), there is an optimal $\phi$ that promotes the cooperation to the largest degree. This observation is in line with previous findings in the single game case that a moderate amount of randomness in small-world networks is found to best enhance cooperation~\cite{ren2007randomness}.

\begin{figure}[t]
\centering
\includegraphics[width=0.99\linewidth]{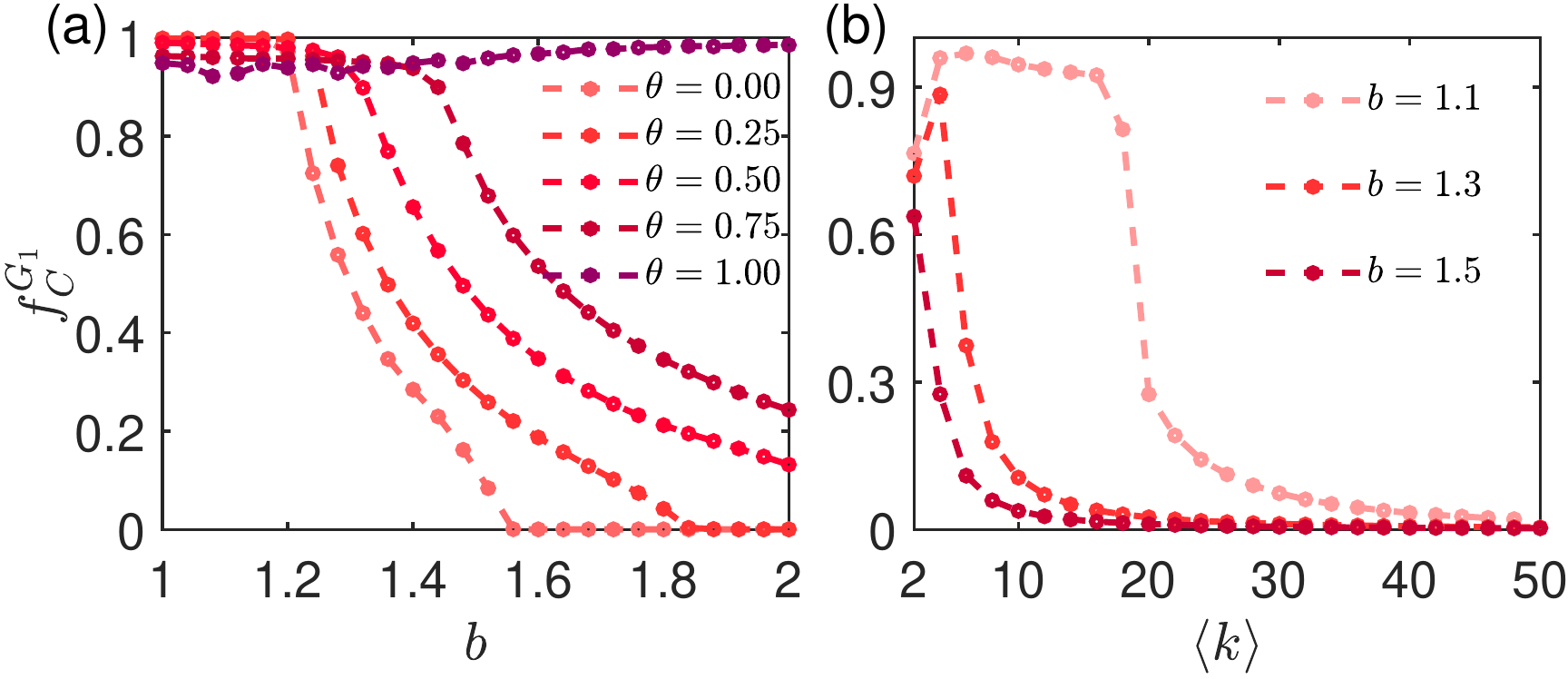}
\caption{(Color online)
Two interacting PD games on Erd\H{o}s-R\'enyi random networks.
(a) The cooperation prevalence $f_c$ versus the temptation $T=b$ for a couple of interaction strengths $\theta$. 
(b) $f_c$ versus the average degree for three temptation values $b$ with $\theta=0.5$.
$f^{G_2}_C\approx f^{G_1}_C$ due to the symmetrical settings.
Parameters: $P=S=0$, $R=1$ for both games, the network size $N=2^{20}$ with $\left\langle k\right\rangle=4$.
}
\label{fig:ER}
\end{figure}

\subsection{Random networks}\label{subsec:RN}
Erd\H{o}s--R\'enyi random networks~\cite{Bollobas2001random} represent a class of topologies found in nature. Its construction starts with an ensemble of $N$ isolated individuals, any two nodes are then connected with a given probability. In such a way, the degrees of the resulting networks follow a Poisson distribution around the mean value $\left\langle k\right\rangle$. 

The evolutionary outcome of two interacting PD games on ER networks is shown in Fig.~\ref{fig:ER}(a). Due to the structural disorder, the prevalence of cooperation is higher than the case of 2d square lattice in the absence of game interaction. By increasing $\theta$, a continuing promotion of cooperation is observed as well, with the cross-playing scenario being the optimal case likewise. Fig.~\ref{fig:ER}(b) shows the dependence of cooperation prevalence on the average degree $\left\langle k\right\rangle$, where there is an optimal $\left\langle k\right\rangle$ for each case, further increasing the value of degree results in a cooperation decline.  This is because the case of $\left\langle k\right\rangle\rightarrow N\!-\!1$ is equivalent to the well-mixed populations, where no cooperation is expected according to the above mean-field analysis. 

\begin{figure}[tbp]
\centering
\includegraphics[width=0.7\linewidth]{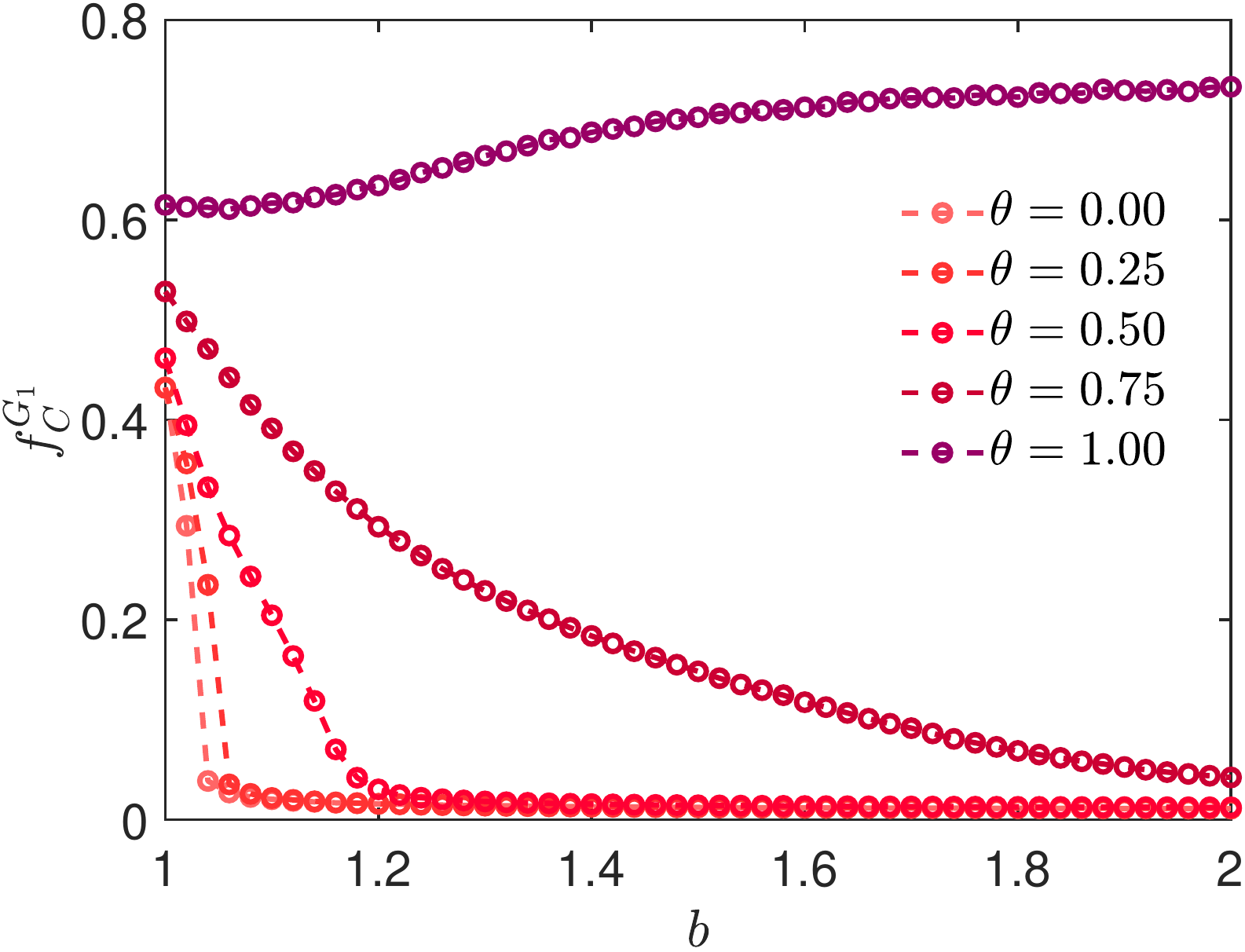}
\caption{(Color online)
Two interacting PD games on BA scale-free networks.
The cooperation prevalence $f_c$ versus the temptation $b$ for a couple of interaction strengths $\theta$. $f^{G_2}_C\approx f^{G_1}_C$ due to the symmetrical settings.
Parameters: $P=S=0$, $R=1$ for both games, the network size $N=2^{20}$ with $\left\langle k\right\rangle=4$. Now $K=0.025$ due to the average payoff scheme, the flipping noise $p_f=0.01$.
}
\label{fig:SF}
\end{figure}

\subsection{Scale-free networks}\label{subsec:SF}
We adopt Barab\'asi--Albert (BA) networks~\cite{Barabasi1999Emergence} for the simulations on the scale-free network. The construction is via the growth and preferential attachment. The network is started with a small fully connected graph as the initial core, and every newly added node is going to connect $\left\langle k\right\rangle /2$ existing nodes, with a probability proportional to their degrees. The generated networks follow a power-law degree distribution with an exponent $-3$.

Different from the SW or ER random networks, which are taken as homogeneous networks, scale-free networks are typical heterogeneous networks. This heterogeneity is able to boost cooperation considerably, yielding fairly high level of cooperation even for a single game. The reason lies in the fact that the hubs are easily to form cooperator backbones that drives the whole network to be cooperative~\cite{Santos2005Scale-Free}. This boost is however based upon the accumulated total payoffs, whereby the hubs are very likely to have higher payoffs and thus be the model players. Once the accumulated payoff is replaced by the average payoff (i.e. $\bar{\Pi}_x=\Pi_x/k_i$), which is argued more reasonable in reality in some previous studies~\cite{Wu2007Evolutionary,Szolnoki2008Towards,Liang2021Social}, the heterogeneity-induced-enhancement is much less significant. Here, we adopt the average payoff scheme in our simulations of two interacting games.

As can be seen in Fig.~\ref{fig:SF}, without game interaction ($\theta=0$), the enhancement effect of heterogeneity is very much inhibited --- a low level of cooperation is seen. As the game interaction $\theta$ increases, the cooperation curve is monotonically lifted, also the cross-playing scenario work best. Note that, in scale-free network case, a flipping noise in posed to inhibited the strong fluctuations caused by the strong degree heterogeneity in the following way: in each elementary step, with a small probability $p_f$ to flip the state of the focal player, and with $1-p_f$ to conduct the standard MC procedure as in Sec. II. With flipping noise, the absorbing state is thus avoided, and it can be interpreted as the deviation from the logic of imitation due to the irrationality.

In brief, additional complexities in the underlying networks of population, as shown in this section, only bring some quantitative difference compare to the lattice case, the dynamical reciprocity still works in the complex networked populations. 

\section{More games}\label{sec:more}
Given the results of two interacting games, one naturally wants to see what's the trend if more games are engaged, since in the real world there are many more issues unfolded simultaneously. The question of interest is: \emph{what could be expected if the number of games involved increases, whether the above mechanism still holds?} 

In~\cite{CSL}, we have shown that by assuming equal contribution for each game, the phase transitions and typical time series show clearly that a higher level of cooperation is expected when more games are involved ($m=1,2,3$). Based upon these observations, one can reasonably extrapolate that a continuing promotion in cooperation is expected when more and more games are engaged. Here, we plan to study the three game case in a bit more depth.
 
The effective payoffs following the linear combination for the three interacting games are now written as 
\begin{equation}
\begin{pmatrix}
\widehat\Pi^{G_1}_{x} \\ \widehat\Pi^{G_2}_{x} \\ \widehat\Pi^{G_3}_{x} 
\end{pmatrix}
\!=\!
\begin{pmatrix}
1\!-\!\theta_2\!-\!\theta_3 & \theta_2 & \theta_3 \\
\theta_1 & 1\!-\!\theta_1\!-\!\theta_3 & \theta_3 \\
\theta_1 & \theta_2 & 1\!-\!\theta_1\!-\!\theta_2  
\end{pmatrix}
\begin{pmatrix}
\Pi^{G_1}_{x} \\ \Pi^{G_2}_{x} \\ \Pi^{G_3}_{x} 
\end{pmatrix},
\label{eq:3game}
\end{equation}
where the interaction strength $\theta_{1,2,3}$ are respectively the contribution weights of $G_{1,2,3}$ in other games' effective payoffs. To reduce the number of parameters, here we adopt a circular parameterization as follows,
\begin{equation}
\begin{pmatrix}
\widehat\Pi^{G_1}_{x} \\ \widehat\Pi^{G_2}_{x} \\ \widehat\Pi^{G_3}_{x} 
\end{pmatrix}
\!=\!
\begin{pmatrix}
1\!-\!\theta'_1\!-\!\theta'_2 & \theta'_1 & \theta'_2 \\
\theta'_2 & 1\!-\!\theta'_1\!-\!\theta'_2 & \theta'_1 \\
\theta'_1 & \theta'_2 & 1\!-\!\theta'_1\!-\!\theta'_2  
\end{pmatrix}
\begin{pmatrix}
\Pi^{G_1}_{x} \\ \Pi^{G_2}_{x} \\ \Pi^{G_3}_{x} 
\end{pmatrix},
\label{eq:3game_reduced}
\end{equation}
where $\theta'_{1,2}\in [0,1]$ and $\theta'_1+\theta'_2\le1$. By varying these two interaction strengths, we can systematically study the case of three interacting games.
 
\begin{figure}[t]
\centering
\includegraphics[width=0.99\linewidth]{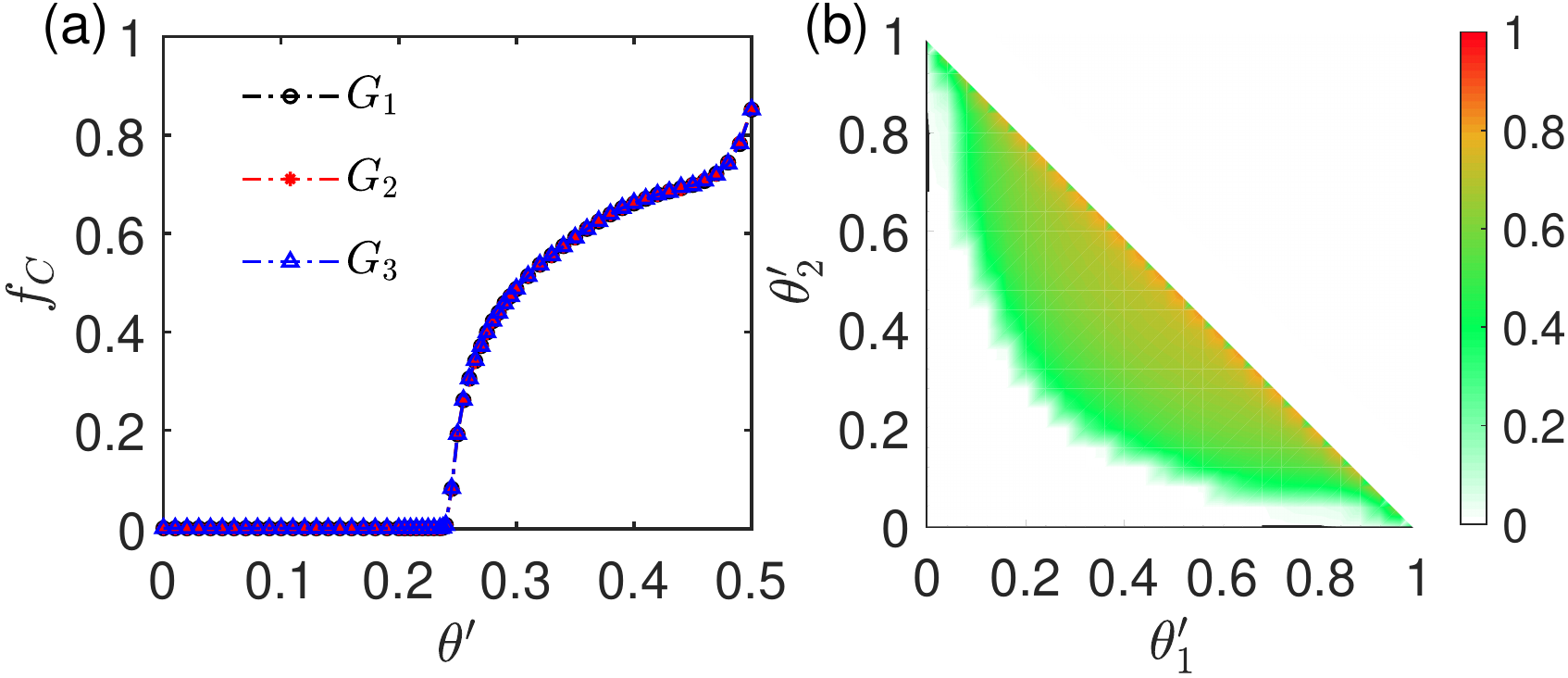}
\caption{(Color online)
Three interacting games.
(a) Phase transitions of the three cooperation prevalences versus the interaction strength $\theta'$ by assuming $\theta'_1=\theta'_2=\theta'$. 
(b) Color-coded fraction of cooperators regarding game $G_1$ ($f^{G_1}_C=\sum_{X,Y\in\mathbb{S}_1}f_{CXY}$) within the interaction space $\theta'_1-\theta'_2$. Due to the restriction of $\theta'_1+\theta'_2\le1$, the upper right half is unphysical. Note that the cooperation prevalence for the three games $f^{G_{1,2,3}}_C$ are approximately identical due to the symmetrical setting, as shown in (a). 
Other parameters: $S=0$, $R=1$, $P=0$, $T=1.1$, $L=1024$ for the 2d square lattice.
}
\label{fig:more}
\end{figure}

The simplest case where $\theta'_1=\!\theta'_2=\!\theta'$ is shown in Fig.~\ref{fig:more}(a). When $\theta'\lesssim 0.25$, cooperators cannot survive within the population for the given parameters, further increasing the interaction strength, a continuous phase transition is seen and the cooperation prevalence goes to be fairly high when $\theta'\rightarrow 1/2$, which corresponds to a cross-playing scenario in the three interacting game case.

Fig.~\ref{fig:more}(b) provides the more general case where $\theta'_{1,2}$ can be different. We see that a stronger game interaction leads to better cooperation holds in general. In particular, high cooperation does not require the symmetry between $\theta'_{1,2}$; in fact, as long as $\theta'_1+\theta'_2\rightarrow 1$, the general cross-playing scenario, fairly high cooperation is guaranteed. 

For three games case, the mechanism behind the promotion is the same as the case of the two interacting games and the above classification still holds. Specifically, there are now 8 different states and 28 pairwise interactions by combination. Even though, these interactions can still be classified into the above three categories also within a cross-playing scenario $\theta'_1\!=\!\theta'_2\!=\!1/2$ for simplicity, as listed in Table~\ref{tab:3games}.

Overall, also two scenarios are considered --- individual and bulk interactions. And all interacting pairs can similarly be classified into invasion, neutral, and catalyzed categories. While no net production in cooperation is expected in the neutral category, cooperators are either yielded or ruined in the other two categories, and the net effect is opposite in two scenarios. Additional complexities here, however, are that for a given pairwise interaction, it could be classified into different categories for different games; e.g.  DCC-CDD belongs to catalyzed type when playing game $G_1$, but is of neutral type when playing game $G_2$ or $G_3$. All classifications are also confirmed by numerical experiments (not shown).

\begin{table*}[]
\begin{tabular}{@{}rlll@{}}
\hline\hline
\toprule
\multicolumn{1}{l}{} & \multicolumn{1}{c}{~~~~~~~~~~~~~~~~~~~~~~~~~Individual interaction~~~~~~~~~~~~~~~~~~~~~~~~~}                                                              & \multicolumn{1}{c}{~~~~~~~~~~~~~~~~~~~~~~~~~~~~Bulk interaction~~~~~~~~~~~~~~~~~~~~~~~~~~~~}                                                                    &  \\ \midrule
\hline
Invasion type~~        & \begin{tabular}[c]{@{}l@{}}
                                  \cellcolor[HTML]{EFEFEF}CCC + DDD $\xrightarrow{  G_1/G_2/G_3  }$ DCC/CDC/CCD + DDD\\     
                                  \cellcolor[HTML]{EFEFEF}CCC + DDC $\xrightarrow{  G_1/G_2  }$ DCC/CDC + DDC\\       
                                  \cellcolor[HTML]{EFEFEF}CCC + CDD $\xrightarrow{  G_2/G_3  }$ CDC/CCD + CDD\\      
                                  \cellcolor[HTML]{EFEFEF}CCC + DCD $\xrightarrow{  G_1/G_3  }$ DCC/CCD + DCD\\      
                                  \cellcolor[HTML]{EFEFEF}DDD + CCD $\xrightarrow{  G_1/G_2  }$ DDD + DCD/CDD\\   
                                  \cellcolor[HTML]{EFEFEF}DDD + DCC $\xrightarrow{  G_2/G_3  }$ DDD + DDC/DCD\\
                                  \cellcolor[HTML]{EFEFEF}DDD + CDC $\xrightarrow{  G_1/G_3 }$ DDD + DDC/CDD\\ \end{tabular} 
                                 & \begin{tabular}[c]{@{}l@{}}
                                  \cellcolor[HTML]{EFEFEF}CCC + DDD $\xrightarrow{  G_2/G_3  }$ CCC + CDD/DCD/DDC\\
                                  \cellcolor[HTML]{EFEFEF}CCC + DDC $\xrightarrow{  G_1/G_2  }$ CCC + CDC/DCC\\       
                                  \cellcolor[HTML]{EFEFEF}CCC + CDD $\xrightarrow{  G_2/G_3  }$ CCC + CCD/CDC\\      
                                  \cellcolor[HTML]{EFEFEF}CCC + DCD $\xrightarrow{  G_1/G_3  }$ CCC + CCD/DCC\\      
                                  \cellcolor[HTML]{EFEFEF}DDD + CCD $\xrightarrow{  G_1/G_2  }$ CDD/DCD + CCD\\   
                                  \cellcolor[HTML]{EFEFEF}DDD + DCC $\xrightarrow { G_2/G_3  }$ DCD/DDC + DCC\\
                                  \cellcolor[HTML]{EFEFEF}DDD + CDC $\xrightarrow{  G_1/G_3  }$ CDD/DDC + CDC\end{tabular}                                                                       &  \\\\

Neutral type~~         & \cellcolor[HTML]{EFEFEF}\begin{tabular}[c]{@{}l@{}}
                                    CCC + DCC $\xrightarrow{    G_1    }$ 2CCC or 2DCC\\ 
                                    CCC + CDC $\xrightarrow{    G_2    }$ 2CCC or 2CDC\\ 
                                    CCC + CCD $\xrightarrow{    G_3    }$ 2CCC or 2CCD\\
                                    DDD + CDD $\xrightarrow{    G_1    }$ 2DDD or 2CDD\\ 
                                    DDD + DCD $\xrightarrow{    G_2    }$ 2DDD or 2DCD\\ 
                                    DDD + DDC $\xrightarrow{    G_3    }$ 2DDD or 2DDC\\ 
                                    CCD + DCD $\xrightarrow{    G_1    }$ 2CCD or 2DCD\\ 
                                    CCD + CDD $\xrightarrow{    G_2    }$ 2CCD or 2CDD\\ 
                                    CDC + CDD $\xrightarrow{    G_3    }$ 2CDC or 2CDD\\ 
                                    CDC + DDC $\xrightarrow{    G_1    }$ 2CDC or 2DDC\\ 
                                    DCC + DDC $\xrightarrow{    G_2    }$ 2DCC or 2DDC\\ 
                                    DCC + DCD $\xrightarrow{    G_3    }$ 2DCC or 2DCD\\ 
                                    CCD + DDC $\xrightarrow{    G_1    }$ CCD + CDC or DCD + DDC\\ 
                                    CCD + DDC $\xrightarrow{    G_2    }$ CCD + DCC or CDD + DDC\\ 
                                    DCC + CDD $\xrightarrow{    G_2    }$ DCC + CCD or DDC + CDD\\ 
                                    DCC + CDD $\xrightarrow{    G_3    }$ DCC + CDC or DCD + CDD\\ 
                                    CDC + DCD $\xrightarrow{    G_1    }$ CDC + CCD or DDC + DCD\\
                                    CDC + DCD $\xrightarrow{    G_3    }$ CDC + DCC or CDD + DCD\\\end{tabular} 
                                    & \cellcolor[HTML]{EFEFEF}\begin{tabular}[c]{@{}l@{}}
                                    CCC + DCC $\xrightarrow{    G_1    }$ 2CCC or 2DCC\\ 
                                    CCC + CDC $\xrightarrow{    G_2    }$ 2CCC or 2CDC\\ 
                                    CCC + CCD $\xrightarrow{    G_3    }$ 2CCC or 2CCD\\
                                    DDD + CDD $\xrightarrow{    G_1    }$ 2DDD or 2CDD\\ 
                                    DDD + DCD $\xrightarrow{    G_2    }$ 2DDD or 2DCD\\ 
                                    DDD + DDC $\xrightarrow{    G_3    }$ 2DDD or 2DDC\\ 
                                    CCD + DCD $\xrightarrow{    G_1    }$ 2CCD or 2DCD\\ 
                                    CCD + CDD $\xrightarrow{    G_2    }$ 2CCD or 2CDD\\ 
                                    CDC + CDD $\xrightarrow{    G_3    }$ 2CDC or 2CDD\\ 
                                    CDC + DDC $\xrightarrow{    G_1    }$ 2CDC or 2DDC\\ 
                                    DCC + DDC $\xrightarrow{    G_2    }$ 2DCC or 2DDC\\ 
                                    DCC + DCD $\xrightarrow{    G_3    }$ 2DCC or 2DCD\\ 
                                    CCD + DDC $\xrightarrow{    G_1    }$ CCD + CDC or DCD + DDC\\ 
                                    CCD + DDC $\xrightarrow{    G_2    }$ CCD + DCC or CDD + DDC\\ 
                                    DCC + CDD $\xrightarrow{    G_2    }$ DCC + CCD or DDC + CDD\\ 
                                    DCC + CDD $\xrightarrow{    G_3    }$ DCC + CDC or DCD + CDD\\ 
                                    CDC + DCD $\xrightarrow{    G_1    }$ CDC + CCD or DDC + DCD\\
                                    CDC + DCD $\xrightarrow{    G_3    }$ CDC + DCC or CDD + DCD\\\end{tabular}  &  \\\\

Catalyzed type~~      & \begin{tabular}[c]{@{}l@{}}
                                   \cellcolor[HTML]{EFEFEF}CDC + DCC $\xrightarrow{   G_1/G_2   }$ CDC/DCC + CCC\\
                                   \cellcolor[HTML]{EFEFEF}CCD + CDC $\xrightarrow{   G_2/G_3   }$ CCD/CDC + CCC\\
                                   \cellcolor[HTML]{EFEFEF}CCD + DCC $\xrightarrow{   G_1/G_3   }$ CCD/DCC + CCC\\
                                   \cellcolor[HTML]{EFEFEF}CDD + DCD $\xrightarrow{   G_1/G_2   }$ CDD/DCD + CCD\\
                                   \cellcolor[HTML]{EFEFEF}DCD + DDC $\xrightarrow{   G_2/G_3   }$ DCD/DDC + DCC\\
                                   \cellcolor[HTML]{EFEFEF}CDD + DDC $\xrightarrow{   G_1/G_3   }$ CDD/DDC + CDC\\
                                   \cellcolor[HTML]{EFEFEF}DCC + CDD $\xrightarrow{   G_1   }$ CCC + CDD\\
                                   \cellcolor[HTML]{EFEFEF}CDC + DCD $\xrightarrow{   G_2   }$ CCC + DCD\\
                                   \cellcolor[HTML]{EFEFEF}CCD + DDC $\xrightarrow{   G_3   }$ CCC + DDC\end{tabular}
                                  & \begin{tabular}[c]{@{}l@{}}
                                   \cellcolor[HTML]{EFEFEF}CDC + DCC $\xrightarrow{   G_1/G_2   }$ DDC + DCC or CDC + DDC\\
                                   \cellcolor[HTML]{EFEFEF}CCD + CDC $\xrightarrow{   G_2/G_3   }$ CDD + CDC or CCD + CDD\\
                                   \cellcolor[HTML]{EFEFEF}CCD + DCC $\xrightarrow{   G_1/G_3   }$ DCD + DCC or CCD + DCD\\
                                   \cellcolor[HTML]{EFEFEF}CDD + DCD $\xrightarrow{   G_1/G_2   }$ DCD/CDD + DDD\\
                                   \cellcolor[HTML]{EFEFEF}DCD + DDC $\xrightarrow{   G_2/G_3   }$ DDC/DCD + DDD\\
                                   \cellcolor[HTML]{EFEFEF}CDD + DDC $\xrightarrow{   G_1/G_3   }$ DDC/CDD + DDD\\
                                   \cellcolor[HTML]{EFEFEF}DCC + CDD $\xrightarrow{   G_1   }$ DCC + DDD\\
                                   \cellcolor[HTML]{EFEFEF}CDC + DCD $\xrightarrow{   G_2   }$ CDC + DDD\\
                                   \cellcolor[HTML]{EFEFEF}CCD + DDC $\xrightarrow{   G_3   }$ CCD + DDD\end{tabular}                                                                       &  \\ \bottomrule
\hline
\end{tabular}
\caption{Categories of interactions in 3 interacting PD games $G_{1,2,3}$ within a cross-playing scheme ($\theta'_1=\theta'_2=1/2$ in Eq.(\ref{eq:3game_reduced})). $C_8^2=28$ pairwise interactions by combination can still be classified into three categories in either individual or bulk scenarios.}
\label{tab:3games}
\end{table*}

\section{Summary and discussions}\label{sec:summary}
In summary, motivated by the facts that different games are often coupled in the real world, we formulate their evolution in the framework of interacting games. We show that this game-game interaction generally boosts the propensities of cooperation in all evolved games. To our surprise, the optimal promotion occurs when the decision-making of the games is completely blind to their own's payoffs. A mean-field treatment reveals that two new routes to cooperation arise, which are confirmed by further analysis. All these findings suggest a new mechanism for cooperation --- dynamical reciprocity. Exhausted numerical evidences for variants both in dynamical and structural aspects have verified the robustness of the dynamical reciprocity. 

While the network reciprocity is to facilitate the growth of cooperator clusters via the structured connectivities~\cite{Nowak2004Evolutionary,szabo2007evolutionary}, the mechanism to lift cooperation in the dynamical reciprocity is through the game-game correlation instead, more specifically through the invasion and catalyzed types of interactions. In this sense, the dynamical reciprocity manifests itself as an entirely new different mechanism. Nonetheless, as shown in the well-mixed case, the interacting games doesn't show any improvement compare to the single game case. This means that the dynamical reciprocity  only works in structured population. Since most populations in the real world are structured, this precondition is easy to meet. Therefore, the two reciprocities are expected to go hand in hand to maintain high levels of cooperation in reality.

In the present work, we have only treated the linear game interaction case as shown Eq.(\ref{eq:effective_linear}). There, the effective payoffs perceived is a linear combination of payoffs for different games, and the interaction strength is encoded within the combination weights. This is of course a simplified formulation for the real cases, more realistic modeling could be more complex functions of the payoffs and the weights. In addition, the effective payoff should also probably incorporate a memory of past history for all evolved games, which implies that a non-Markovian model~\cite{wang2006memory,Kampen2007stochastic} seems more proper.

The five strategy updating rules used in this work essentially belong to the outward learning, where they imitate those who are better off. This is the mainstream approach of modeling updating. However, a new paradigm proposed recently~\cite{zhang2020oscillatory,zhang2020understanding} is through the inward learning, where the decision-making of individuals is through introspective actions based on their histories. With the help of machine learning~\cite{carleo2019machine}, they also model the evolution of cooperation, but is limited to the single game case. It would be interesting to see what if more games are played in this shifted paradigm, does the dynamical reciprocity still work? 

As the next step, maybe the most important open question, is to verify the dynamical reciprocity in behavioral experiments. But due to the great complexities in human beings, the experiment needs to be carefully designed to be convincing. Ideally, the game-game correlation is tunable; also the comparison of interacting games within structured and unstructured populations is necessary according to our results.

Back to the real world, the implications of our works is at least two facets. On one hand, since different issues are often interweaved in the real world, and highly cooperative behaviors are abundant, the dynamical reciprocity seemingly provides a natural causality explanation for these two observations. On the other hand, to handle those cooperation failures in some vital issues, such as global warming and trade war, our work suggests that the players should get involved in as many games as possible, whereby a decent cooperation should be expected as a result. This advice could be applicable to different contexts from international affairs to interpersonal relationships.

Finally, our work may also provide an inspiration for the complexity science. In complexity science, many systems are studied within the framework of structure plus dynamics (i.e. the function), where the structure and its impact on dynamics have been extensively studied with the rise of network science. Our findings suggest that the dynamics-dynamics interactions may also harbor a great amount of complexities, which have largely been underestimated before. Another good example of ``more is different"~\cite{anderson1972more} in dynamics is modeling the spread of multiple infectious diseases~\cite{Cai2015Avalanche,Peter2016Phase,Chen2017Fundamental,Chen2019Persistent}, where the physics revealed is fundamentally different from the single one case. We hope the related communities could put more attention to the potential complexities arising from the dynamics-dynamics interactions in the future.


\textit{Acknowledgements ---}
This work is supported by the National Natural Science Foundation of China under Grant Nos 61703257 and 12075144, and by the Fundamental Research Funds for the Central Universities GK201903012. 
L. C. acknowledges the enlightening discussions with Dirk Brockmann (HU and RKI) in the early phase of the project, and Ying-Cheng Lai (ASU) for helpful comments.

\appendix
\begin{widetext}
\section{Detailed mean-field treatment}\label{app:A}
\setcounter{equation}{0}
\renewcommand{\theequation}{A\arabic{equation}}
Following Eq.(\ref{eq:MF_RE}) in Sec. IV, we explicitly write down the \emph{effective fitness} defined as $\widehat{\Pi}_{s}^{G_{1,2}}=(1-\theta)\Pi_{s}^{G_{1,2}}+\theta\Pi_{s}^{G_{2,1}}, s\in\mathbb{S}_2$ that players perceived and used in their strategy updating, as follows
\begin{equation}
     \left\{
                \begin{array}{ll}
                \widehat{\Pi}_{CC}^{G_1}=(1-\theta)(f_{CC}R+f_{CD}R+f_{DC}S+f_{DD}S)+\theta(f_{CC}R+f_{CD}S+f_{DC}R+f_{DD}S),\\
		\widehat{\Pi}_{CC}^{G_2}=(1-\theta)(f_{CC}R+f_{CD}S+f_{DC}R+f_{DD}S)+\theta(f_{CC}R+f_{CD}R+f_{DC}S+f_{DD}S),\\
		\widehat{\Pi}_{CD}^{G_1}=(1-\theta)(f_{CC}R+f_{CD}R+f_{DC}S+f_{DD}S)+\theta (f_{CC}T+f_{CD}P+ f_{DC}T+ f_{DD}P),\\
		\widehat{\Pi}_{CD}^{G_2}=(1-\theta)(f_{CC}T+f_{CD}P+f_{DC}T+f_{DD}P)+\theta (f_{CC}R+f_{CD}R+ f_{DC}S+ f_{DD}S),\\
		\widehat{\Pi}_{DC}^{G_1}=(1-\theta)(f_{CC}T+f_{CD}T+f_{DC}P+f_{DD}P)+\theta (f_{CC}R+f_{CD}S+ f_{DC}R+f_{DD}S),\\
		\widehat{\Pi}_{DC}^{G_2}=(1-\theta)(f_{CC}R+f_{CD}S+f_{DC}R+f_{DD}S)+\theta (f_{CC}T+ f_{CD}T+ f_{DC}P+f_{DD}P),\\
		\widehat{\Pi}_{DD}^{G_1}=(1-\theta)(f_{CC}T+f_{CD}T+f_{DC}P+f_{DD}P)+\theta(f_{CC}T+f_{CD}P+f_{DC}T+f_{DD}P),\\
		\widehat{\Pi}_{DD}^{G_2}=(1-\theta)(f_{CC}T+f_{CD}P+f_{DC}T+f_{DD}P)+\theta(f_{CC}T+f_{CD}T+f_{DC}P+f_{DD}P).
                \end{array}
        \right. 
\end{equation}

The overall effective payoffs are then
\begin{equation}
\widehat{\Pi}_{XY}=\widehat{\Pi}^{G_1}_{XY}+\widehat{\Pi}^{G_2}_{XY}=(1-\theta)\Pi^{G_1}_{XY}+\theta\Pi^{G_2}_{XY}+
(1-\theta)\Pi^{G_2}_{XY}+\theta\Pi^{G_1}_{XY}=\Pi^{G_1}_{XY}+\Pi^{G_2}_{XY}=\Pi_{XY}. 
\end{equation}\label{eq:pi_overall}
The key observation here is that the interaction strength $\theta$ is cancelled out due to the linear combination, which is also reasonable since $\theta$ is the contribution weight that only adjusts the payoff values perceived in a specific game but not the overall payoffs.
Specifically, we have

\begin{equation}
     \left\{
                \begin{array}{ll}
                \widehat{\Pi}_{CC}=(2f_{CC}+f_{CD}+f_{DC})R+(2f_{DD}+f_{CD}+f_{DC})S,\\ 
		\widehat{\Pi}_{CD}=(f_{CC}+f_{CD})R+(f_{DC}+f_{DD})S+(f_{CC}+f_{DC})T+(f_{CD}+f_{DD})P,\\ 
		\widehat{\Pi}_{DC}=(f_{CC}+f_{CD})T+(f_{DC}+f_{DD})P+(f_{CC}+f_{DC})R+(f_{CD}+f_{DD})S,\\ 
		\widehat{\Pi}_{DD}=(2f_{CC}+f_{CD}+f_{DC})T+(2f_{DD}+f_{CD}+f_{DC})P.
                \end{array}
        \right. 
\end{equation}

And the mean fitness is
\begin{equation}
\bar{\Pi}=\sum_{s\in\mathbb{S}_2}f_{s}\widehat{\Pi}_{s}=f_{CC}\widehat{\Pi}_{CC}+f_{CD}\widehat{\Pi}_{CD}+f_{DC}\widehat{\Pi}_{DC}+f_{DD}\widehat{\Pi}_{DD}.\label{eq:MF_meanf}
\end{equation}

Inset these terms into Eq. (\ref{eq:MF_RE}), the replicator equations are then 
\begin{equation}
     \left\{
                \begin{array}{ll}
                \dot{{f}}_{CC}=f_{CC}[f_{CD}(\widehat{\Pi}_{CC}-\widehat{\Pi}_{CD}) + f_{DC}(\widehat{\Pi}_{CC}-\widehat{\Pi}_{DC}) + f_{DD}(\widehat{\Pi}_{CC}-\widehat{\Pi}_{DD})], \\ 
		\dot{{f}}_{CD}=f_{CD}[f_{CC}(\widehat{\Pi}_{CD}-\widehat{\Pi}_{CC}) + f_{DC}(\widehat{\Pi}_{CD}-\widehat{\Pi}_{DC}) + f_{DD}(\widehat{\Pi}_{CD}-\widehat{\Pi}_{DD})], \\
		\dot{{f}}_{DC}=f_{DC}[f_{CC}(\widehat{\Pi}_{DC}-\widehat{\Pi}_{CC}) + f_{CD}(\widehat{\Pi}_{DC}-\widehat{\Pi}_{CD}) + f_{DD}(\widehat{\Pi}_{DC}-\widehat{\Pi}_{DD})], \\
		\dot{{f}}_{DD}=f_{DD}[f_{CC}(\widehat{\Pi}_{DD}-\widehat{\Pi}_{CC}) + f_{CD}(\widehat{\Pi}_{DD}-\widehat{\Pi}_{CD}) + f_{DC}(\widehat{\Pi}_{DD}-\widehat{\Pi}_{DC})],
                \end{array}
        \right. \label{eq:2MF}
\end{equation}
where we used the normalization condition $f_{CC}\!+\!f_{CD}\!+\!f_{DC}\!+\!f_{DD}\!=\!1$. These equations can actually be summarized as 
\begin{equation}
\dot{{f}}_{s}=\sum_{s'\in\mathbb{S}_2}[f_{s}f_{s'}(\widehat{\Pi}_{s}-\widehat{\Pi}_{s'}) ].\label{eq:MF_brief}
\end{equation}
The structure of Eq. (\ref{eq:MF_brief}) is similar to Eq. (\ref{eq:MF_single}) and its meaning is straightforward that the change in $f_s$ comes from the interaction of individuals within state $s$ and $s'$ and their effective payoff difference. The overall effective fitness differences are as follows
\begin{equation}
     \left\{
                \begin{array}{ll}
                 \widehat{\Pi}_{CC}-\widehat{\Pi}_{CD} =  (f_{CC}+f_{DC})(R-T) + (f_{CD}+f_{DD})(S-P),\\ 
		\widehat{\Pi}_{CC}-\widehat{\Pi}_{DC} = (f_{CC}+f_{CD})(R-T) + (f_{DD}+f_{DC})(S-P),\\
		\widehat{\Pi}_{CC}-\widehat{\Pi}_{DD} = (2f_{CC}+f_{CD}+f_{DC})(R-T) + (2f_{DD}+f_{CD}+f_{DC})(S-P),\\
		\widehat{\Pi}_{CD}-\widehat{\Pi}_{DC} = (f_{CD}-f_{DC})(R-T) + (f_{DC}-f_{CD})(S-P),\\
		\widehat{\Pi}_{CD}-\widehat{\Pi}_{DD} = (f_{CC}+f_{CD})(R-T) + (f_{DC}+f_{DD})(S-P),\\
		\widehat{\Pi}_{DC}-\widehat{\Pi}_{DD} = (f_{CC}+f_{DC})(R-T) + (f_{CD}+f_{DD})(S-P).
                \end{array}
        \right. 
\end{equation}
Now let us consider the evolution of overall cooperator fraction with regard to $G_1$ by adding the first two equations in Eqs.~(\ref{eq:2MF}) and insert the above related terms
\small
\begin{eqnarray*}
\dot{{f}}^{G_1}_{C}&=&\dot{f}_{CC}+\dot{f}_{CD}\\
                             & = &f_{CC}\{f_{DC}(\widehat{\Pi}_{CC}-\widehat{\Pi}_{DC})+f_{DD}(\widehat{\Pi}_{CC}-\widehat{\Pi}_{DD})\}+f_{CD}\{f_{DC}(\widehat{\Pi}_{CD}-\widehat{\Pi}_{DC})+f_{DD}(\widehat{\Pi}_{CD}-\widehat{\Pi}_{DD})\}\\
                             & = & f_{CC}\{f_{DC}[(f_{CC}+f_{CD})(R-T)+(f_{DC}+f_{DD})(S-P)]  
                                    + f_{DD}[(2f_{CC}+f_{CD}+f_{DC})(R-T) + (2f_{DD}+f_{CD}+f_{DC})(S-P)]\} \\
                             &    & + f_{CD}\{f_{DC}[(f_{CD}-f_{DC})(R-T) + (f_{DC}-f_{CD})(S-P)]
                                    + f_{DD}[(f_{CC}+f_{CD})(R-T) + (f_{DC}+f_{DD})(S-P)]\}\\
                             & = & (f_{CC}f_{DC}+f_{CD}f_{DD})[(f_{CC}+f_{CD})(R-T)+(f_{DC}+f_{DD})(S-P)]  
                                      + f_{CC}f_{DD}[(2f_{CC}+f_{CD}+f_{DC})(R-T)\\ 
                             &    & + (2f_{DD}+f_{CD}+f_{DC})(S-P)] 
                                     + f_{CD}f_{DC}[(f_{CD}-f_{DC})(R-T) + (f_{DC}-f_{CD})(S-P)].
\end{eqnarray*}
\normalsize
By replacing $f_{CC}f_{DC}\!+\!f_{CD}f_{DD}\!=\!(f_{CC}\!+\!f_{CD})(f_{DC}\!+\!f_{DD})\!-\!(f_{CC}f_{DD}\!+\!f_{CD}f_{DC})$,
\small
\begin{eqnarray*}                             
\dot{{f}}^{G_1}_{C}& = & (f_{CC}+f_{CD})(f_{DC}+f_{DD})[(f_{CC}+f_{CD})(R-T)+(f_{DC}+f_{DD})(S-P)] \\ 
                             &    & + f_{CC}f_{DD}\{[(2f_{CC}+f_{CD}+f_{DC})(R-T) + (2f_{DD}+f_{CD}+f_{DC})(S-P)] 
                                       - [(f_{CC}+f_{CD})(R-T)+(f_{DC}+f_{DD})(S-P)]\}\\
                             &    & + f_{CD}f_{DC}\{[(f_{CD}-f_{DC})(R-T) + (f_{DC}-f_{CD})(S-P)]
                                       - [(f_{CC}+f_{CD})(R-T)+(f_{DC}+f_{DD})(S-P)]\}\\
                             & = & f^{G_1}_{C}f^{G_1}_{D}[f^{G_1}_C(R-T) + f^{G_1}_D(S-P)]
                                       + (f_{CC}f_{DD}-f_{CD}f_{DC})[f^{G_2}_C(R-T) + f^{G_2}_D(S-P)]\\
                             & = & f^{G_1}_{C}f^{G_1}_{D}(\Pi^{G_1}_C-\Pi^{G_1}_D) + (f_{CC}f_{DD}-f_{CD}f_{DC})(\Pi^{G_2}_C-\Pi^{G_2}_D).
\end{eqnarray*}
\normalsize

Similarly, one can also obtain the equation for $G_2$ by adding the first and the third equations in Eqs.~(\ref{eq:2MF}).
\begin{eqnarray}
\dot{{f}}^{G_2}_{C}&=&\dot{f}_{CC}+\dot{f}_{DC} \nonumber \\
                             & = & f^{G_2}_{C}f^{G_2}_{D}(\Pi^{G_2}_C-\Pi^{G_2}_D) + (f_{CC}f_{DD}-f_{CD}f_{DC})(\Pi^{G_1}_C-\Pi^{G_1}_D).\label{2MF2}
\end{eqnarray}
Equation (\ref{eq:mf_cor}) is then obtained. 

\section{Analytical solutions}\label{app:B}
\setcounter{equation}{0}
\renewcommand{\theequation}{B\arabic{equation}}
To solve the mean-field equation Eq. (\ref{eq:2MF}), we make some rearrangements that lead to

\begin{equation}
     \left\{
                \begin{array}{ll}
                  \dot{f}_{CC} =  2f_{CC}f^{G_1}_{D}[f^{G_1}_{C}(R-T)+f^{G_1}_{D}(S-P)],\\
                  \dot{f}_{CD} = f_{CD}(-f^{G_1}_{C}+f^{G_1}_{D})[f^{G_1}_{C}(R-T)+f^{G_1}_{D}(S-P)], \\
                  \dot{f}_{DC} = f_{DC}(-f^{G_1}_{C}+f^{G_1}_{D})[f^{G_1}_{C}(R-T)+f^{G_1}_{D}(S-P)],\\
                  \dot{f}_{DD} = -2f_{DD}f^{G_1}_{C}[f^{G_1}_{C}(R-T)+f^{G_1}_{D}(S-P)],
                \end{array}
              \right. 
\end{equation}
where $f^{G_1}_C=f_{CD}+f_{CC}$ and $f^{G_1}_D=f_{DC}+f_{DD}$ by definition.  Based on the observations in the numerical simulation as well as the symmetrical game setting, it's reasonable to assume $f^g_{CD}=f^g_{DC}$, $g\in \mathbb{G}$.
Together with the normalization condition, only two  variables (let's select $f_{CC}$ and $f_{DD}$) are independent, the others can be expressed by
\begin{eqnarray}
 f_{CD}&=&f_{DC}=\dfrac{1-f_{CC}-f_{DD}}{2},\\
 f^{G_1}_{C}&=&\dfrac{1+f_{CC}-f_{DD}}{2},\\
 f^{G_1}_{D}&=&\dfrac{1-f_{CC}+f_{DD}}{2}.
\end{eqnarray}
Their equations are
\begin{equation}
     \left\{
                \begin{array}{ll}
                \dot{f}_{CC}=f_{CC}(1-f_{CC}+f_{DD})[\dfrac{(1+f_{CC}-f_{DD})}{2}(R-T)+\dfrac{(1-f_{CC}+f_{DD})}{2}(S-P)],\\
                \dot{f}_{DD}=-f_{DD}(1+f_{CC}-f_{DD})[\dfrac{(1+f_{CC}-f_{DD})}{2}(R-T)+\dfrac{(1-f_{CC}+f_{DD})}{2}(S-P)].
                \end{array}
              \right. 
\end{equation}
 
By setting $\dot{f}_{CC}=\dot{f}_{DD}=0$, we obtain the four solutions:
\begin{eqnarray}
(1) &\quad & f_{CC}=f_{DD}=0; \label{eq:MF_s1}\\
(2) &\quad & f_{CC}=1, f_{DD}=0; \label{eq:MF_s2}\\
(3) &\quad & f_{CC}=0, f_{DD}=1; \label{eq:MF_s3}\\
(4) &\quad & \dfrac{(1+f_{CC}-f_{DD})}{2}(R-T)+\dfrac{(1-f_{CC}+f_{DD})}{2}(S-P)=0. \label{eq:MF_s4}
\end{eqnarray}
The stability of these solutions is determined by the eigenvalues of the corresponding Jacobian matrix as
\begin{equation}
J=\begin{pmatrix}
\dfrac{\partial \dot{f}_{CC}}{\partial f_{CC}} & \dfrac{\partial \dot{f}_{CC}}{\partial f_{DD}} \\
\\
\dfrac{\partial \dot{f}_{DD}}{\partial f_{CC}} & \dfrac{\partial \dot{f}_{DD}}{\partial f_{DD}} 
\end{pmatrix}.
\end{equation}

(1) For $f_{CC}=f_{DD}=0$,
\small
\begin{equation}
J=\begin{pmatrix}
\dfrac{(R+S-T-P)}{2} & 0 \\
\\
0&  -\dfrac{(R+S-T-P)}{2}
\end{pmatrix}.
\end{equation}
\normalsize
We have $\lambda=\pm\dfrac{(R+S-T-P)}{2}$, the eigenvalues couldn't be both negative, thus this solution is unstable in any case.

(2) For $f_{CC}=1,f_{DD}=0$,
\small
\begin{equation}
J=\begin{pmatrix}
-(R-T) & (R-T) \\
\\
0& -2(R-T)
\end{pmatrix}.
\end{equation}
\normalsize
We have $\lambda_{1}=-(R-T),\lambda_{2}=-2(R-T)$. Therefore, this solution is stable only when $T<R$.

(3) For $f_{CC}=0,f_{DD}=1$,
\small
\begin{equation}
J=\begin{pmatrix}
2(S-P) & 0 \\
\\
-(S-P)& (S-P)
\end{pmatrix}.
\end{equation}
\normalsize
We have $\lambda_{1}=(S-P),\lambda_{2}=2(S-P)$. This solution is stable only when $S<P$.

(4) For $\dfrac{(1+f_{CC}-f_{DD})}{2}(R-T)+\dfrac{(1-f_{CC}+f_{DD})}{2}(S-P)=0$,  or equivalently $f^{G_1}_C=\dfrac{P-S}{R+P-S-T}$,
\small
\begin{equation}
J=\begin{pmatrix}
f_{CC}(1-f_{CC}+f_{DD})\dfrac{(R-S-T+P)}{2} & -f_{CC}(1-f_{CC}+f_{DD})\dfrac{(R-S-T+P)}{2} \\
\\-f_{DD}(1+f_{CC}-f_{DD})\dfrac{(R-S-T+P)}{2}& f_{DD}(1+f_{CC}-f_{DD})\dfrac{(R-S-T+P)}{2}
\end{pmatrix}.
\end{equation}
\normalsize
We have $\lambda_{1}=0,\lambda_{2}=(f_{CC}+f_{DD}-f^{2}_{CC}-f^2_{DD})\dfrac{(R-S-T+P)}{2}$. This solution is stable only when $(R-S-T+P)<0$, since $f_{CC}+f_{DD}-f^{2}_{CC}-f^2_{DD}>0$.

Put together, the stable solutions of the mean-field equations are exactly recovered to the single pairwise game: the mixed state (solution (\ref{eq:MF_s4})) is stable for SD game region, full cooperation (solution (\ref{eq:MF_s2})) is stable for HG game region, full defection (solution (\ref{eq:MF_s3})) is stable for PD game region, and SH region is bistable (full cooperation or full defection) because it is the overlapped region for solution (\ref{eq:MF_s2}) and (\ref{eq:MF_s3}).
\end{widetext}
\bibliography{manuscript}
\end{document}